\documentclass[structabstract]{aa}
\usepackage{graphicx}
\usepackage{amsmath}
\usepackage{txfonts}
\newcommand{\Msol}{M_\odot}
\newcommand{\degree}{^\circ}
\newcommand{\gsct}{\gamma \ {\rm Sct} }
\newcommand{\gnor}{\gamma \ {\rm Nor} }
\newcommand{\bsct}{\beta \ {\rm Sct} }
\newcommand{\tmus}{\theta \ {\rm Mus} }
\newcommand{\ud}{{\mathrm d}}

\begin{document} 
\title{Understanding EROS2 observations
toward the spiral arms within a classical Galactic model framework}
\author{
M.~Moniez\inst{1},
S.~Sajadian\inst{2},
M.~Karami\inst{3}$^{,}$\inst{4},
S.~Rahvar\inst{5},
R.~Ansari\inst{1},
}
\institute{
Laboratoire de l'Acc\'{e}l\'{e}rateur Lin\'{e}aire,
{\sc IN2P3-CNRS}, Universit\'e de Paris-Sud, B.P. 34, 91898 Orsay Cedex, France
\and
Department of Physics, Isfahan University of Technology, Isfahan 84156-83111, Iran
\and
Perimeter Institute for Theoretical Physics, 31 Caroline Street North, Waterloo,
Ontario N2L 2Y5, Canada
\and
Department of Physics and Astronomy, University of Waterloo,
200 University Avenue West, Waterloo, ON N2L 3G1, Canada
\and
Department of Physics, Sharif University of Technology, P.O. Box 11155-9161, Tehran, Iran
}

\offprints{M. Moniez, \email{ moniez@lal.in2p3.fr} \\
{\it see also our WWW server at  URL :} \\
{\tt http://www.lal.in2p3.fr/recherche/eros}}

\date{Received ??/??/2009, accepted }
%

\abstract
{}{
EROS (Exp\'erience de Recherche d'Objets Sombres) has searched for microlensing toward four directions in the Galactic plane away from the Galactic center. The interpretation of
the catalog optical depth is complicated by the spread of the
source distance distribution. We compare the EROS microlensing observations
with Galactic models (including the Besan\c{c}on model), tuned to fit
the EROS source catalogs, and
take into account all observational data such as the microlensing optical
depth, the Einstein crossing durations, and the  
color and magnitude distributions of the catalogued stars.
\\
}
{
We simulated EROS-like source catalogs using the
HIPPARCOS (HIgh-Precision PARallax COllecting Satellite) database, the Galactic mass distribution,
and an interstellar extinction table. Taking into account the EROS
star detection
efficiency, we were able to produce simulated color-magnitude diagrams
that fit the observed diagrams. This allows us to estimate average microlensing
optical depths and event durations that are directly comparable with the measured values.\\
}
{
Both the Besan\c{c}on model and our Galactic model allow us to fully understand the EROS
color-magnitude data. The average optical depths and mean event durations calculated from these
models are in reasonable agreement with the observations.
Varying the Galactic structure parameters through simulation, we were also able to deduce contraints
on the kinematics of the disk, the disk stellar mass function (at a few $kpc$ distance from the Sun),
and the maximum contribution of a thick disk of compact objects in
the Galactic plane ($M_{thick}< 5-7\times 10^{10} \Msol$ at $95\%$, depending on the model).
We also show that the microlensing data toward one of our monitored directions are significantly
sensitive to the Galactic bar parameters, although much larger statistics are needed to provide
competitive constraints.
\\
}
{
Our simulation gives a better understanding of the lens and source spatial
distributions in the microlensing events. The goodness of a global fit
taking into account all the observables (from the color-magnitude diagrams and
microlensing observations) shows the validity of the
Galactic models. Our tests with the parameters excursions show the unique
sensitivity of the microlensing data to the kinematical parameters and stellar initial mass function (IMF).
}

\titlerunning{Microlensing toward the spiral arms}
\authorrunning{Moniez, Sajadian, Karami, Rahvar, Ansari}
\maketitle

\section{Introduction}

\begin{figure}[htbp]
\begin{center}
\includegraphics[width=9cm]{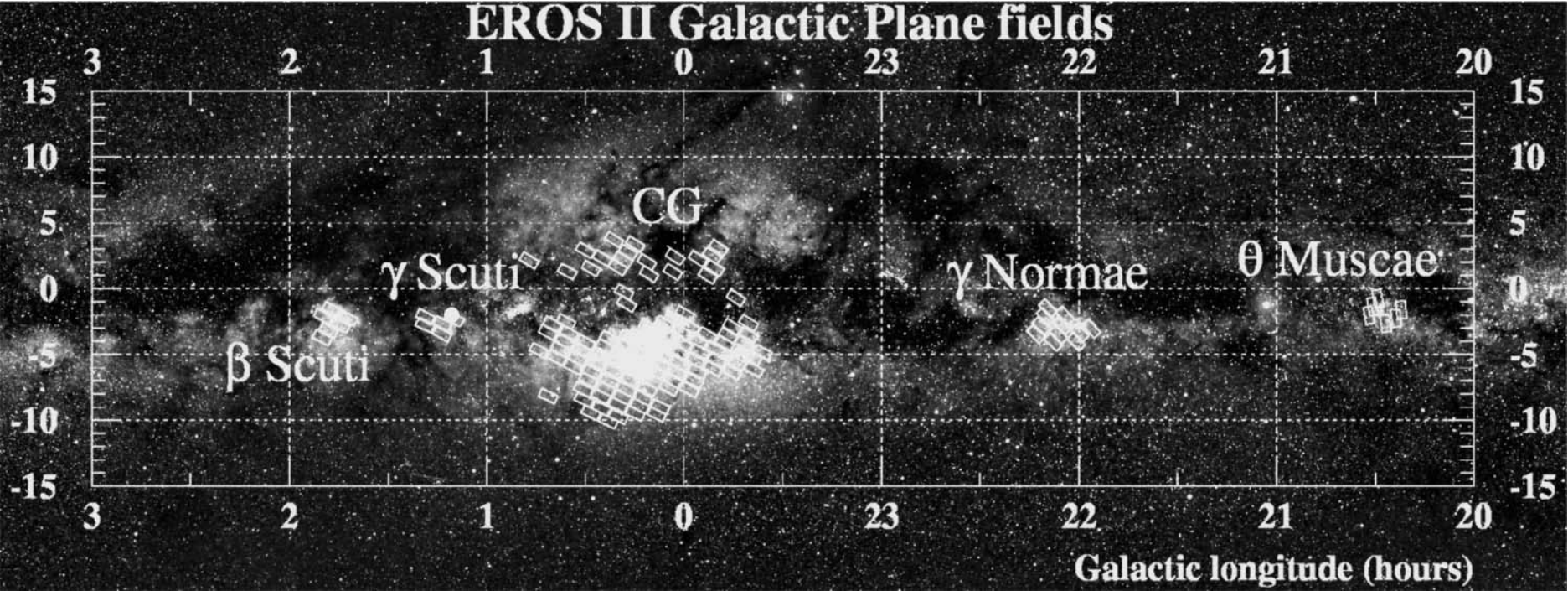}
\includegraphics[width=7cm]{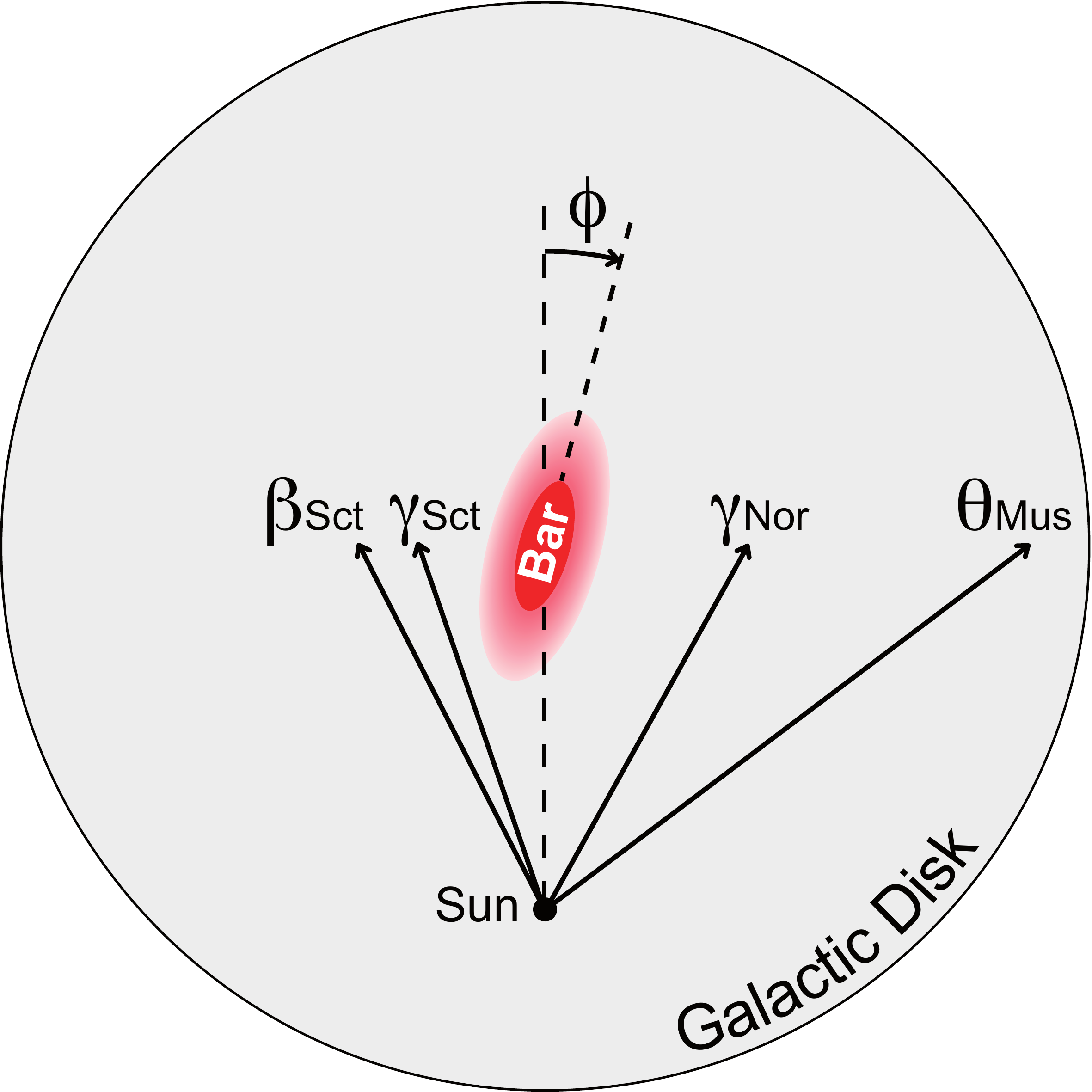}
\caption[]{
Four directions toward the Galactic spiral arms monitored by EROS.
}
\label{BSfields}
\end{center}
\end{figure}

Following Paczy\'nskis' seminal publication (\cite{pacz1986}),
several groups initiated survey programs beginning in 1989
to search for compact halo objects within the Galactic halo.
The challenge for the EROS (Exp\'erience de Recherche d'Objets Sombres)
and MACHO (MAssive Compact Halo Objects)
teams was to clarify the status of the missing hadrons in our own Galaxy.
In September 1993, the three teams, EROS (\cite{eroslmc}),
MACHO (\cite{machlmc}), and
OGLE (Optical Gravitational Lensing Experiment, \cite{oglpr}), discovered the first microlensing
events in the directions of the Large Magellanic Cloud and
the Galactic center (GC). 
Since these first discoveries, thousands of microlensing effects have been
detected in the direction of the GC together with
a handful of events toward the Galactic spiral arms (GSA) and the
Magellanic Clouds. \\
Microlensing has proven to be a powerful probe of the
Milky Way structure.
Searches for microlensing toward the Magellanic Clouds (LMC, SMC)
and M31 (survey MEGA ; \cite{Crotts}
and survey AGAPE; \cite{M31}) provide optical depths through the Galactic halo,
allowing one to study dark matter in the form of massive compact objects.
Searches toward the Galactic plane (GC and Galactic
spiral arms) allow one to measure the microlensing optical depth of ordinary
stars in the Galactic disk and bar.
Kinematical models and mass functions can also be constrained through
the event duration distributions. \\
Several teams have published
results about the Galactic structure, through microlensing searches in the Galactic plane,
such as MACHO (\cite{machobul2005}), EROS (\cite{Hamadache}), OGLE (\cite{Sumi2006}),
and MOA (Microlensing Observations in Astrophysics, \cite{Awiphan}).
The EROS team is the only group that have searched for microlensing toward the
Galactic spiral arms, away from the Galactic center.
As a matter of fact, the EROS team have measured the microlensing optical depth toward
four directions of the Galactic plane (Fig. \ref{BSfields}), {\it i.e.},
\begin{itemize}
\item
$\gsct\ (\bar b=-2.1\degree,\bar l=18.5\degree)$,
\item
$\gnor\ (-2.4\degree, 331.1\degree)$,
\item
$\bsct\ (-2.2\degree, 26.6\degree)$,
\item
$\tmus\ (-1.5\degree, 306.6\degree)$,
\end{itemize}
as far as 55 degrees in longitude away from the Galactic center
(\cite{BS7ans}).
The specificity of these measurements with respect to other
targets like SMC or LMC is the widespread distribution of the
distances of the monitored sources. The distances to the sources could not
be individually measured and both their average and dispersion
are poorly estimated.
The concept of ``catalog optical depth'' was introduced in \cite{BS7ans},
and in this paper we describe a complete procedure to compare
measured optical depth with model predictions.
After the introduction of the microlensing concepts (Sect. \ref{section:basics}) and
the presentation of the EROS data (Sect. \ref{section:EROSdata}),
in Sect. \ref{section:generation} we describe the technique to produce synthetic 
color-magnitude diagrams (CMDs), via the HIPPARCOS catalog (HIgh-Precision PARallax COllecting Satellite
\cite{hipp}, \cite{Turon}),
the spatial distribution of mass from Galactic models,
and the absorptions tabulated in a 3D map obtained with infrared observations (\cite{Marshall}).
We cross-checked the obtained local stellar number densities with the
expectations from the stellar initial mass function (IMF).
In Sect. \ref{section:comparison}, we describe the full simulation of the EROS program, in terms
of CMDs taking into account the stellar detection efficiency of EROS, and in terms of the microlensing events. 
Our fitting procedure is described in Sect. \ref{section:fitting}, where we derive
constraints on our simple Galactic model and test the Besan\c{c}on model (\cite{besancon});
the fit takes into account the observed CMDs as well as the data from the microlensing
(optical depths and mean event durations) toward the four observed lines of sight;
we use the fit to estimate the allowed range of our simple Galactic model parameters.
In the final discussion (Sect. \ref{section:discussion}), we extract from the best fit the distance
distributions of the sources and lenses.
Finally, we discuss the sensitivity of microlensing observations toward the Galactic arms
to the dark thick disk, central bar inclination, stellar mass function and disk kinematics.
\section{Microlensing effect}
\label{section:basics}
The gravitational microlensing effect occurs when a massive compact object
passes close enough to the line of sight of a star
to produce a temporary magnification of the source.
A general overview of the microlensing formalism can be found in \cite{Schneider} and \cite{Rahvar}.
In the approximation of a single point-like
lens deflecting the light from a single point-like source,
the total magnification of the source luminosity
at time $t$ is given by (\cite{pacz1986})
\begin{equation}
\label{magnification}
A(t)=\frac{u(t)^2+2}{u(t)\sqrt{u(t)^2+4}}\ ,
\end{equation}
where $u(t)$ is the distance of the deflecting object
to the undeflected line of sight, expressed
in units of the Einstein radius $R_{\mathrm{E}}$ given by:
\begin{eqnarray}
R_{\mathrm{E}} &=& \sqrt{\frac{4GM}{c^2}D_S x(1-x)} \\
&\simeq& 4.54\ \mathrm{A.U.} \times\left[\frac{M}{\Msol}\right]^{\frac{1}{2}}
\left[\frac{D_S}{10\ kpc}\right]^{\frac{1}{2}}
\frac{\left[x(1-x)\right]^{\frac{1}{2}}}{0.5}. \nonumber
\end{eqnarray}
Here $G$ is the Newtonian gravitational constant,
$D_S$ is the distance of the observer to the source, and
$x D_S=D_L$ is its distance to the deflector of mass $M$.
Assuming a deflector moving at a constant relative transverse
speed $v_T$, reaching its minimum
distance $u_0$ (impact parameter) to the undeflected line of sight
at time $t_0$, $u(t)$ is given by
\begin{equation}
\label{impact}
u(t)=\sqrt{u_0^2+\left( \frac{t-t_0}{t_{\mathrm{E}}}\right)^2},
\end{equation}
where $t_{\mathrm{E}}=R_{\mathrm{E}} /v_T$, the lensing timescale, is
the only measurable parameter
bringing useful information regarding the lens parameters in the
approximation of simple microlensing,
\begin{eqnarray}
t_{\mathrm{E}} \sim
79\ \mathrm{days} \times 
\left[\frac{v_T}{100\, km/s}\right]^{-1}
\left[\frac{M}{\Msol}\right]^{\frac{1}{2}}
\left[\frac{D_S}{10\, kpc}\right]^{\frac{1}{2}}
\frac{[x(1-x)]^{\frac{1}{2}}}{0.5}\; . 
\end{eqnarray}
\subsection{Microlensing event characteristics}
The so-called simple microlensing effect (point-like source and
point-like lens
with uniform relative motion with respect to the line of sight)
has some characteristic
features that allow one to discriminate it from any known
intrinsic stellar variability. These features are as follows:
given the low probability for source detector alignment within
$R_E$, the event should be singular in the history of the source
(as well as of the deflector);
the 
magnification is independent of the color;
the magnification is a simple function of
time, depending on ($u_0, t_0, t_{\mathrm{E}}$),
with a symmetrical shape;
as the geometric configuration of the source-deflector system
is random, the impact parameters of the events must be uniformly distributed;
the passive role of the lensed stars implies that their
population should be representative of the monitored sample
at any given source distance, particularly with respect to the observed color and
magnitude distributions.

This simple microlensing description can be complicated in
many different ways: for example, multiple lens and source systems (\cite{Mao}),
extended sources (\cite{Yoo}), and parallax effects (\cite{Gould}); these complications
will not be discussed here.
\subsection{Observables: optical depth, event rate, and $t_{\mathrm{E}}$ distribution}
The optical depth up to a given source distance, $D_S$, is defined as the
instantaneous probability for the line of sight of a target source to intercept
a deflector's Einstein disk, which corresponds to a magnification $A > 1.34$.
Assuming that the distribution of the deflector masses is
described by a density function $\rho(D_L)$ and a
normalized mass function $\ud n_L(D_L,M)/\ud M$, this probability is
\begin{equation}
\tau(D_S)=\int_0^{D_S}\int_{M=0}^{\infty} \frac{\pi \theta_E^2}{4\pi} \times
\frac{\rho(D_L)}{M}\frac{\ud n_L(D_L,M)}{\ud M}\ud M 4\pi D_L^2\ud D_L\, ,
\end{equation}
where $\theta_E=R_{\mathrm{E}}/D_L$ is the angular Einstein radius of a lens of mass
$M$ located at $D_L$.
The second term of the integral is the differential number of these lenses
per mass unit.
As the solid angle of the Einstein disk is proportional to
the deflectors' mass $M$, this probability is found to be independent of
the deflectors' mass function
\begin{equation}
\tau(D_S)=\frac{4 \pi G D_S^2}{c^2}\int_0^1 x(1-x)\rho(x) \ud x\, ,
\end{equation}
where $\rho(x)$ is the mass density of deflectors located at a
distance $x D_S$.
This expression is used when the distance to the monitored
source population is known (for example, toward the LMC and SMC).

When the monitored population is spread over a wide distance distribution,
as is the case toward the Galactic plane, we have to consider the
concept of ``catalog optical depth'' as introduced in \cite{BS7ans};
the mean optical depth toward a given population
defined by a distance distribution $\ud n_S(D_S)/\ud D_S$
of target stars is defined as (\cite{Moniez})
\begin{equation}
<\tau>
=\frac{\int_0^{\infty} \frac{\ud n_S(D_S)}{\ud D_S} \tau(D_S)D_S^2\, \ud D_S}
{\int_0^{\infty} \frac{\ud n_S(D_S)}{\ud D_S}  D_S^2\, \ud D_S}.
\label{optpop}
\end{equation}
Again, this optical depth does not depend on the deflectors' mass
function.
On the other hand, for a given optical depth, the microlensing event rate
depends on the deflectors' mass distribution as well as on the
velocity and spatial distributions.

Contrary to the optical depth, the microlensing event durations $t_E$ and consequently
the event rate (deduced from the optical depth and durations) depend on the deflectors’ mass distribution
as well as on the velocity and spatial distributions.
The statistical properties of the durations and event rates can therefore provide global information on the
dynamics of the Galaxy and on the mass distribution, which complement other observational techniques
based on direct velocity and luminosity measurements.

In this paper, the optical depth together with the observed event rate and more precisely the duration distributions
are compared with simulations to constrain the mass, shape and kinematics of the lensing structures.

\section{EROS data toward the Galactic spiral arms}
\label{section:EROSdata}
\begin{figure}[htbp]
\begin{center}
\includegraphics[width=9cm]{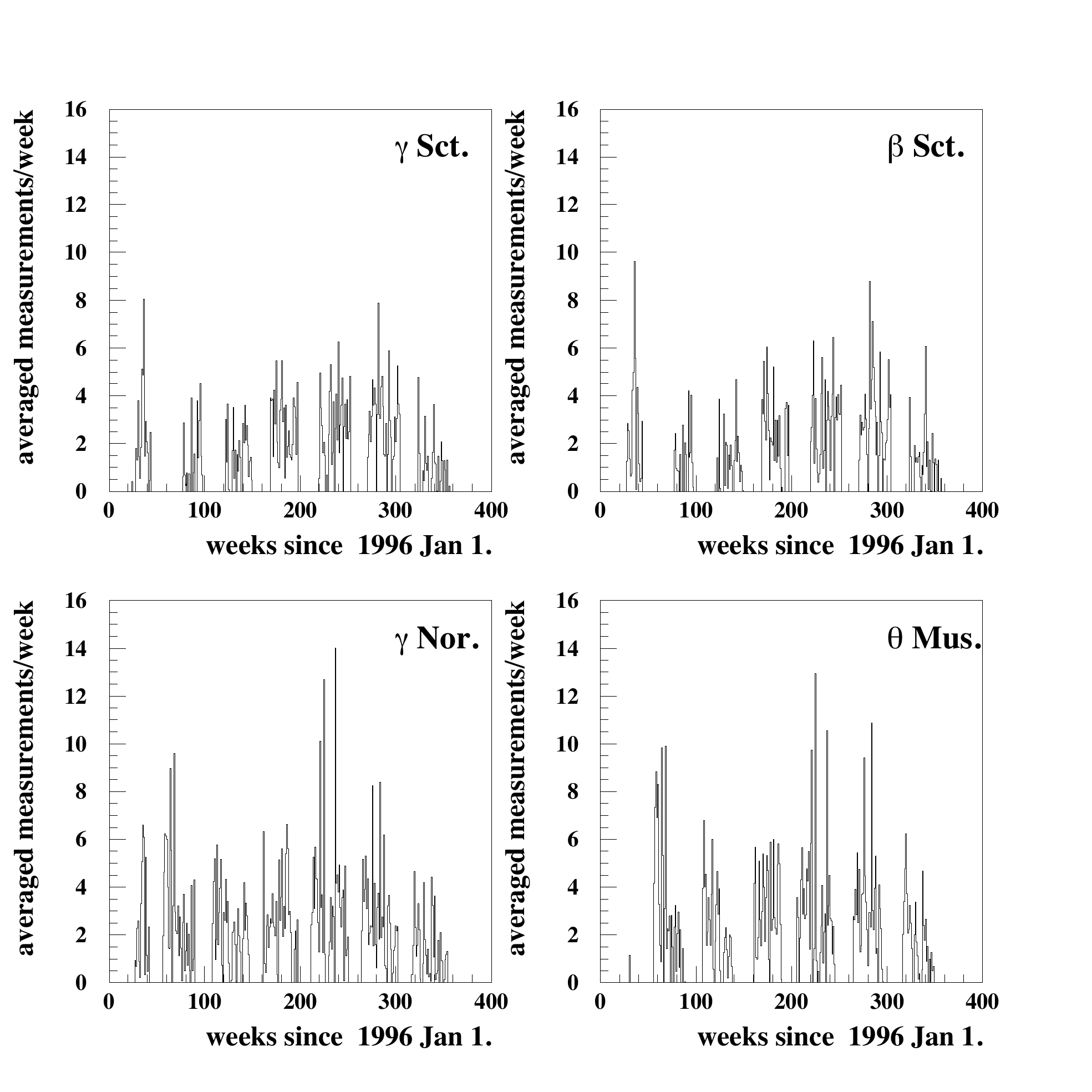}
\caption[]{
Time sampling toward the 4 monitored targets in the Galactic spiral arms:
average number of measurements per star and per week.
}
\label{sampling}
\end{center}
\end{figure}

In this section, we recall and summarize the
EROS2 CCD observations and microlensing results 
toward the Galactic spiral arms,
and describe the efficiencies and uncertainties needed to allow comparisons with simulations.
Fig. \ref{sampling} shows the observation time span with the average weekly sampling toward
the four targets discussed here. We only provide the information on the data that is relevant
for our simulation; more details on the original data can be found in (\cite{BS7ans}).

\subsection{EROS color-magnitude diagrams}
\begin{figure}[htbp]
\begin{center}
\includegraphics[width=9cm]{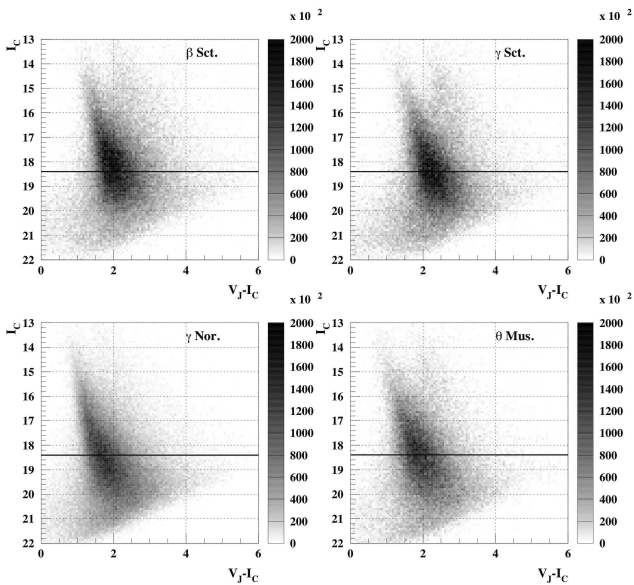}
\caption[]{
Relative color-magnitude diagrams $n(I,V-I)$
of the EROS catalogs toward the 4 directions toward the
Galactic spiral arms.
The gray scale gives the number density of stars per square degree,
unit of magnitude, and unit of color index.
}
\label{HRdiagrams}
\end{center}
\end{figure}

The stars detected in EROS are statistically described by their
color-magnitude diagrams given in Fig. \ref{HRdiagrams} in the $(I_C,V_J)$
photometric system, hereafter simply noted $(I,V)$.
The published EROS-CMDs provide for each catalog, labeled $(C)$,
the observed stellar density $n_C(I,V-I)$ 
per square degree, magnitude, and color index, as a function
of $I$ and $V-I$, sampled in $0.3\times 0.2$ cells (\cite{BSweb}).
When using these CMDs, one has to take into account the following uncertainties:
\begin{itemize}
\item
Each stellar number density $n_C(I,V-I)$ value is affected by 
a statistical uncertainty coming from the propagation of the Poissonian noise
in the original EROS catalogs, as explained in the header of the published EROS-CMD (\cite{BSweb}).
\item
Each $n_C(I,V-I)$ value is affected by 
a systematic uncertainty of $\sim 5.3\%$, owing to the uncertainty
on the size of the effective EROS field; this uncertainty is common to
all catalogs.
\item
Another systematic uncertainty is due to the residual
$0.07$ magnitude EROS calibration uncertainty (\cite{Blanc}),
which affects the attribution of a star to a given $[I,(V-I)]$ cell.
It has to be taken into account for each EROS color, and therefore induces a systematic uncertainty
of $[0.07,0.16] mag$. in the $[I,V-I]\equiv [R_{EROS},(B_{EROS}-R_{EROS})/0.6]$ system.
\end{itemize}

\begin{figure}[htbp]
\begin{center}
\includegraphics[width=9.5cm]{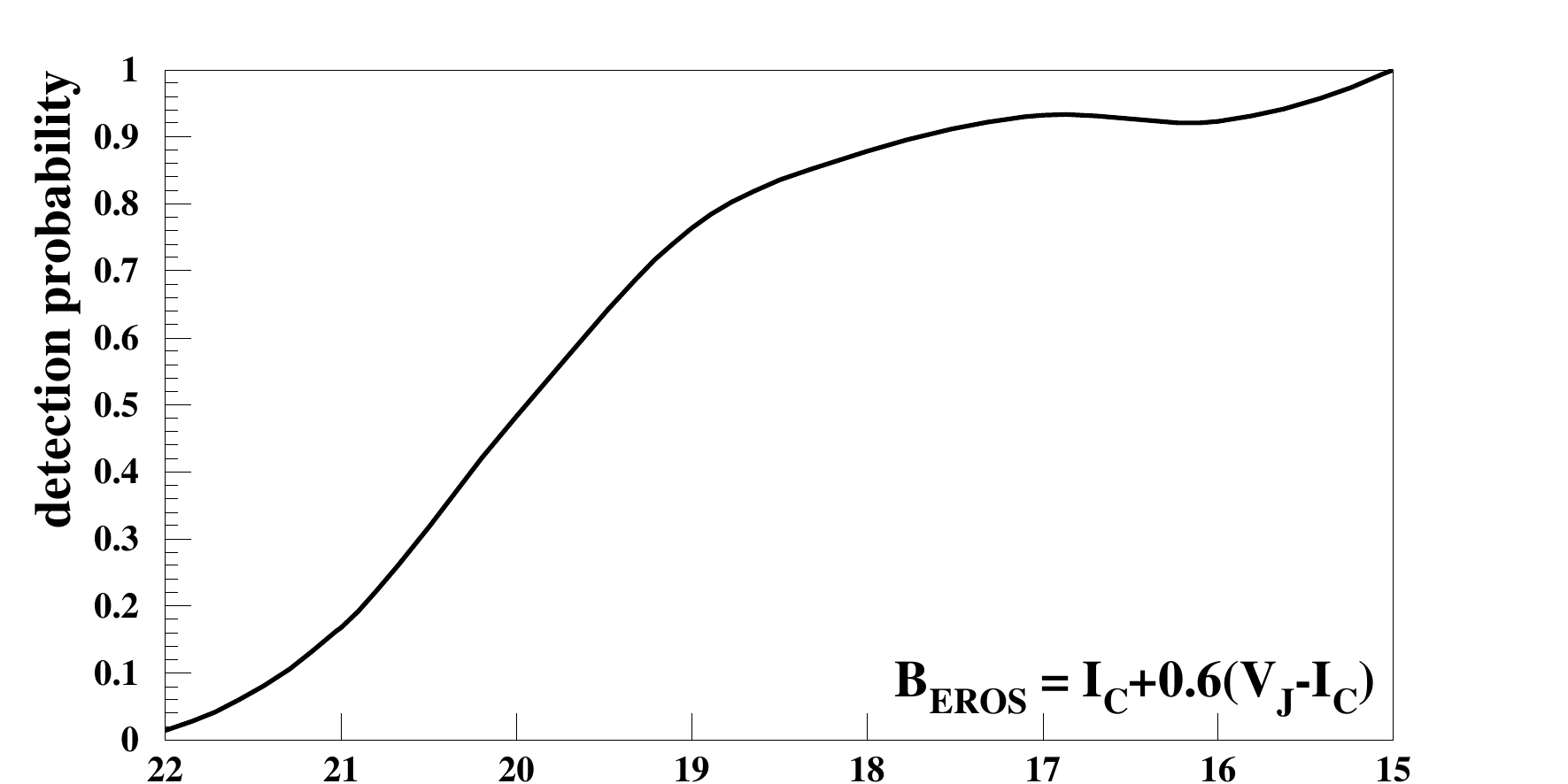}
\caption[]{
Star detection probability in EROS vs. the relative magnitude
$B_{EROS}=I+0.6(V-I)$.
}
\label{effdet}
\end{center}
\end{figure}

\begin{figure}[htbp]
\begin{center}
\includegraphics[width=9.5cm]{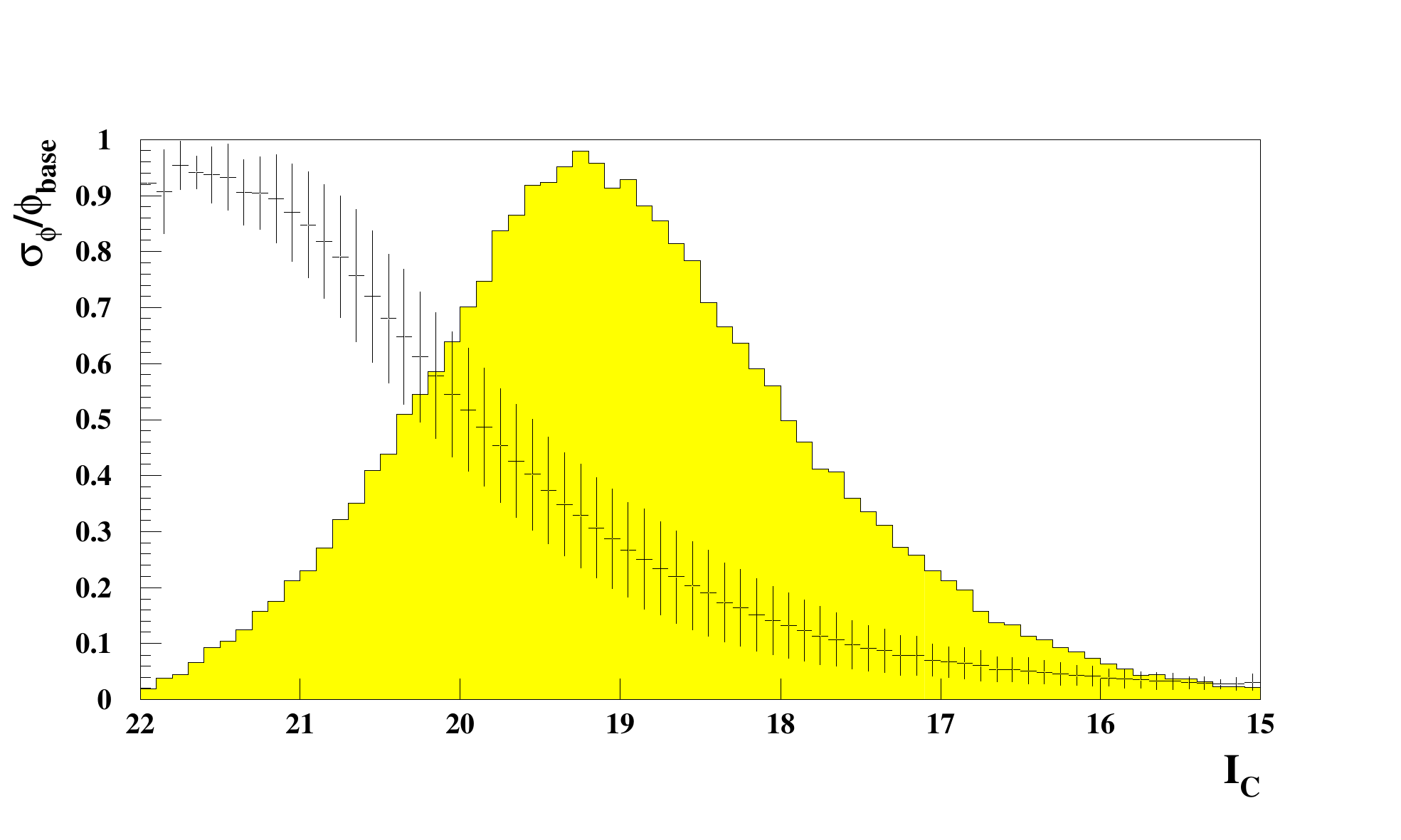}
\includegraphics[width=8.5cm]{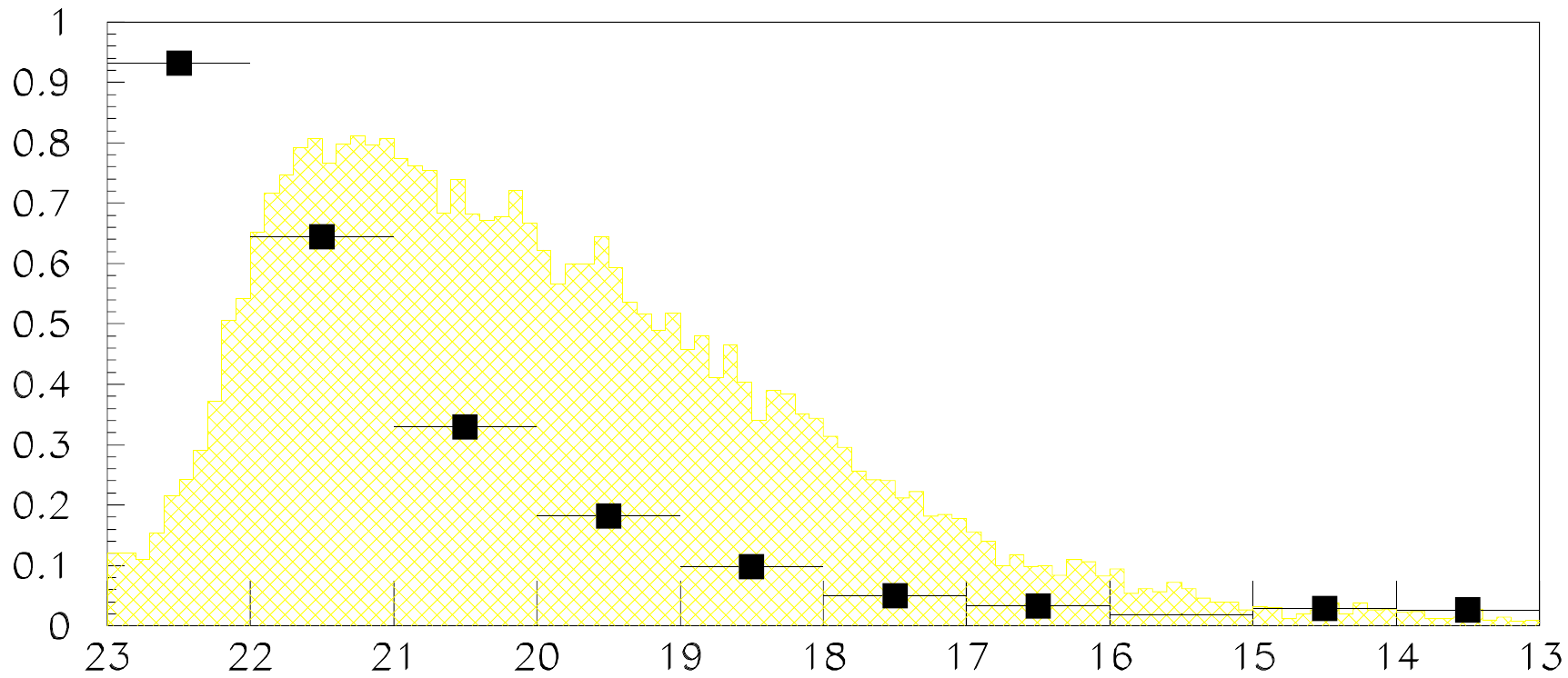}
\caption[]{
Photometric point-to-point precision along the EROS light-curves vs.
$R_{EROS}=I$ (upper) and $B_{EROS}$ (lower). Vertical bars in $I$
show the dispersion of this
precision in the EROS catalog. The histograms show the magnitude
distribution of the full EROS spiral arm catalog (all directions).
}
\label{resolution}
\end{center}
\end{figure}

To generate an ``EROS-like'' catalog from a model for comparison puroposes,
one needs to use the efficiency of EROS to detect stars
and the photometric uncertainties, both defined in the EROS
photometric system $[R_{EROS},B_{EROS}]\equiv [I,I+0.6(V-I)]$.
The EROS stellar detection efficiency has been studied
in (\cite{BS7ans}), by comparing EROS data with HST data
(\cite{HSTarchive}). Since we found that an object detected in
$B_{EROS}$ is systematically detected in $R_{EROS}$, the EROS
stellar detection efficiency can be parametrized as a function of the relative
magnitude $B_{EROS}$ only (Fig. \ref{effdet}).
The EROS photometric errors on the magnitudes and colors are parametrized as
\begin{equation}
\delta I=\sqrt{0.1^2+
\left[\frac{2.5}{\ln
    10}\right]^2\left[\frac{\sigma_{\Phi}}{\Phi}\right]_{R_{EROS}}^2\frac{1}{N_{meas.}}},
\label{photoerr}
\end{equation}
\begin{equation}
\label{photoerrcol}
\delta(V-I)\!\!=\!\! 
\sqrt{0.1^2\!+\!\left[\frac{1}{0.6}\frac{2.5}{\ln
    10}\right]^2\left(\left[\frac{\sigma_{\Phi}}{\Phi}\right]_{R_{EROS}}^2\!\!\!
  +\!\left[\frac{\sigma_{\Phi}}{\Phi}\right]_{B_{EROS}}^2
\right)\frac{1}{N_{meas.}}}, \nonumber
\end{equation}
where the $0.1$ constant term (dominant for stars brighter than $\sim 18$)
is a residual uncertainty, as estimated from EROS calibration
studies using DENIS catalog data (\cite{DENIS})\footnote{This irreducible uncertainty
is attributed to the variability of the stellar spectra within the very wide EROS passbands.},
$\left[\sigma_{\Phi}/\Phi\right]$ is the relative image-to-image dispersion of the successive
flux measurements given by Fig. \ref{resolution}, and
$N_{meas}$ is the number of observations (exposures) used to estimate the mean flux of
a star, {\it i.e.},
268 toward $\beta$ Sct,
277 toward $\gamma$ Sct,
454 toward $\gamma$ Nor and
375 toward $\theta$ Mus.

Table \ref{tabfields} summarizes some of the key numbers regarding the color-magnitude
statistical data. When comparing the data with simulations, we focus on the stars brighter than $I=18.4$,
the most reliable part of the EROS-CMD, with the highest and best controlled stellar detection efficiency.

\subsection{Microlensing results}
Table \ref{tabfields} provides the microlensing results from EROS (\cite{BS7ans}).
The $\sigma_{t_E}$ values differ from the values published in table 3 from (\cite{BS7ans})
because they were biased, since we assumed large statistics for their estimates.
To properly take into account the statistical fluctuations on small numbers,
we therefore re-estimated $\sigma_{t_E}$ from expression,
\begin{equation}
\sigma^2_{t_E} = \frac{1}{N_{events}-1}\sum_{events}(t_E-\overline{t_E})^2,
\end{equation}
where $N_{events}$ is the number of microlensing events toward the target.

\begin{figure}[htbp]
\begin{center}
\includegraphics[width=9cm]{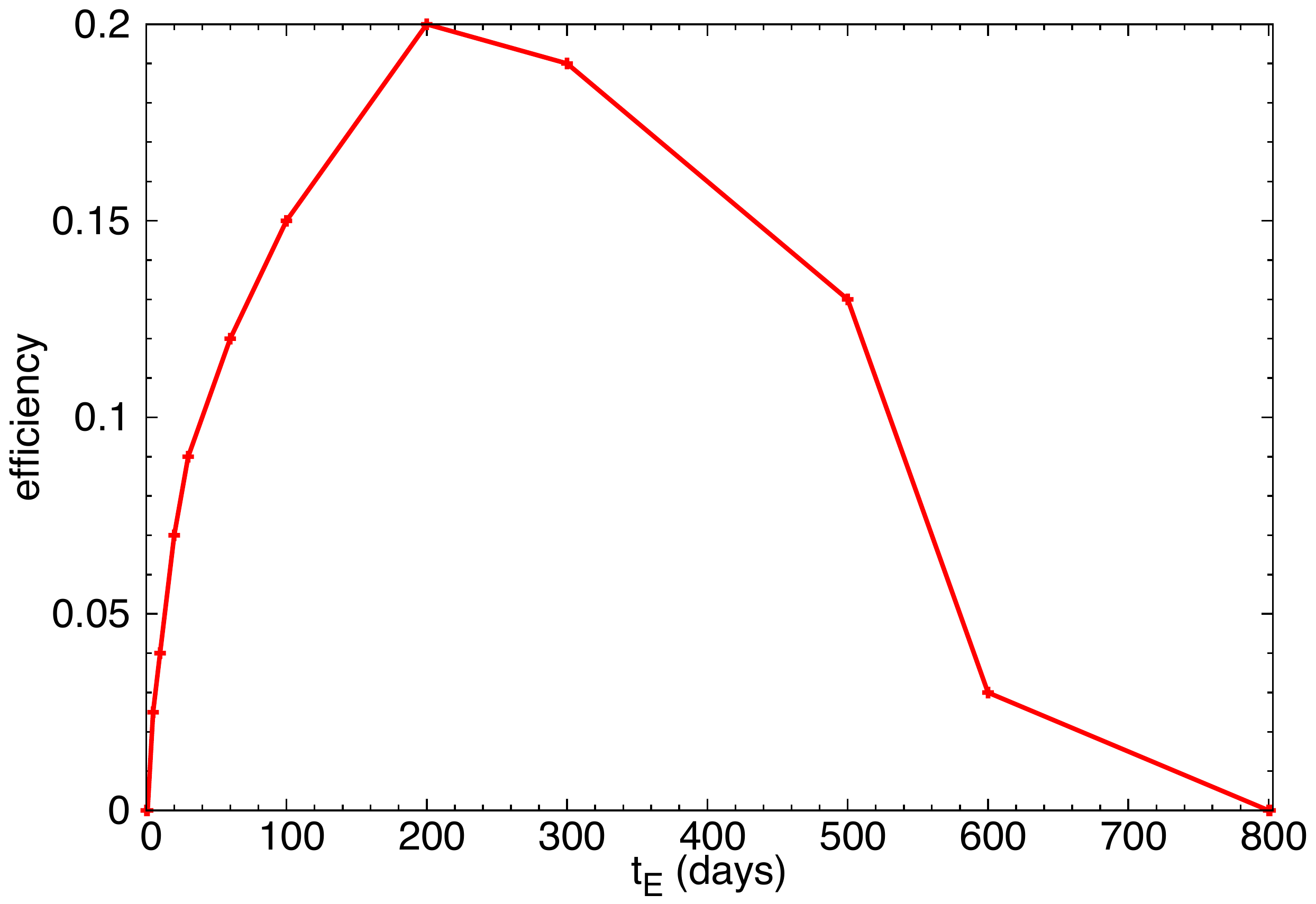}
\caption[]{
Microlensing detection efficiency of the EROS survey toward the Galactic spiral arms,
as a function of the event characteristic duration $t_E$.
}
\label{efficiency}
\end{center}
\end{figure}

The average microlensing detection efficiency of the EROS survey was estimated in \cite{BS7ans};
it is defined as the ratio of events satisfying the EROS selection cuts to the
theoretical number of events with an impact parameter $u_0 < 1$,
and was found to be almost independent of the target, since the time samplings were
very similar.
Figure \ref{efficiency} shows this efficiency as a function of the Einstein duration of the
events $t_E$.

\begin{table}[h!] 
\begin{center}
\caption{ Data and results toward the 4 regions monitored in the EROS spiral arms program.
Average coordinates, field extensions, numbers of bright stars ($I<18.4$),
surface densities of all stars, of the bright stars, and the measured microlensing optical
depth and duration parameters are provided for each target.}
\label{tabfields}
{\scriptsize
\begin{tabular}{c c c c c}\hline
Target & $\tmus$ & $\gnor$ & $\gsct$ & $\bsct$  \\
\hline
$<\alpha\degree>$ & 200 & 245 & 278 & 281 \\
$<\delta\degree>$ & -64 & -52 & -13 & -6 \\
$<b\degree >$     & -1.46 & -2.42 & -2.09 & -2.15 \\ 
$<l\degree >$     & 306.56 & 331.09 & 18.51 & 26.60 \\
\hline
field ($\mathrm{deg^2}$) & 3.8 & 8.4 & 3.6 & 4.3 \\
\hline
$N_{stars}^{I_c<18.4}$ & $2.28\, 10^6$ & $5.24\, 10^6$ & $2.38\, 10^6$
& $3.0\, 10^6$ \\
\hline
$\rho_{*} \times 10^{-6} \mathrm{deg^{-2}}$ 	& 0.60 & 0.62 & 0.66 & 0.70\\
\hline
$\rho_{*}^{I<18.4} \times 10^{-6} \mathrm{deg^{-2}}$  & 0.245	& 0.23	& 0.28	& 0.34 \\
\hline
$N_{event}$ with $u_0<0.7$ & 3 & 10 & 6 & 3 \\
$\tau\times 10^6$    &$.67^{+.63}_{-.52}$&$.49^{+.21}_{-.18}$&$.72^{+.41}_{-.28}$&$.30^{+.23}_{-.20}$ \\ 
\hline
$\overline{t_E}$ (days) & $97\pm 75$ & $57\pm 10$ & $47\pm 6$ & $59\pm 9$ \\
$\sigma_{t_E}$ (days) & 98 & 31 & 14 & 12 \\
\hline
\end{tabular}
}
\end{center}
\end{table}

\section{How to synthesize an EROS-like color-magnitude diagram}
\label{section:generation}
We now compare the data with realistic simulations.
In this section we describe how our modeling takes into account all the known
observational constraints and discuss how to handle the specific difficulties of this kind of analysis.

We generated apparent color-magnitude diagrams based on the following
hypotheses and ingredients from direct observations:
\begin{itemize}
\item
The HIPPARCOS catalog (\cite{hipp}, \cite{Turon}) provides the magnitudes
and colors of 118218 local stars. We assume that the local population
is representative of the entire Galactic disk stellar population. This hypothesis
is certainly justified for the disk stars. The central bar stellar
population is redder, but the EROS observations we are considering here
do not point toward its center.
\item
A random magnitude shift is induced to take into account
observational limitations, such as blending and uncertainties, from
the HIPPARCOS and EROS data.
\item
The spatial mass density distribution results from the
addition of the contributions of thin and thick disks and of the bar modeled according
to \cite{Binney} and \cite{Dwek} or to the Besan\c{c}on model (\cite{besancon}).
\item
The light propagation is affected by Galactic extinction in $I$ and reddening in $V-I$, obtained
from a 3D table of $K_S$ extinctions kindly provided by (\cite{Marshalltable}).
\end{itemize}

\subsection{Producing a CMD from the local HIPPARCOS catalog}
\label{section:unbias}
We present in Appendix A our procedure to obtain a debiased CMD in the $(I,V)$ color system
within the domain $0 < M_V < 8$ from the HIPPARCOS catalog. This debiased catalog
is described by the distribution $n({\bf M})$, where ${\bf M}$ represents
the absolute magnitude and color ``vector'' of a given stellar type.
We established in Appendix A that the numerical contribution of stars brighter
than $M_V=0$ is negligible in a deep Galactic image. In our case, given our
limiting magnitude, we can also neglect the contribution of stars fainter than $M_V=8$.

Assuming that the stellar composition is constant along the line of sight,
stars of any given type are distributed along the
line proportionally to the total mass density $\rho$.
The number of stars expected per square degree
($\Omega (1\degree\!\!\times\! 1\degree)= 3.046\times 10^{-4}\, sr$)
in the EROS catalog is then the integral along the line of sight
\begin{eqnarray}
&&n_{EROS} ({\bf m}) =  \label{catcont} \\
&&\int_{0}^{\infty} \!\!\!\frac{\rho(D)}{\rho_\odot}
n({\bf m}-\delta{\bf m}-{\bf \mu}(D)-{\bf A}(D))  \epsilon_{EROS} ({\bf m})\Omega
(1\degree\!\!\times\! 1\degree) D^2 \ud D, \nonumber
\end{eqnarray}
where
\begin{itemize}
\item
$D$ is the distance to the star along the line of sight,
\item
$\mu(D)$ the corresponding distance modulus (independent on the color),
\item
${\bf A}(D)$ is the interstellar extinction vector (one component per filter)
\item
$\delta{\bf m}$ is a random shift of ${\bf m}$ that takes into account blending (see Sect. \ref{section:blending})
and uncertainties from HIPPARCOS parallax and EROS photometry;
HIPPARCOS stellar absolute $I$ magnitudes are randomly shifted
according to a Gaussian distribution of dispersion
\begin{equation}
\epsilon_I=\sqrt{\left[5 \log e \times\frac{\delta\pi}{\pi}\right]^2+(\delta I)^2},
\label{errorI}
\end{equation}
where $\pi$ and $\delta\pi$ are the HIPPARCOS parallax and associated error, and
$\delta I$ is the estimated EROS photometric uncertainty from expression (\ref{photoerr}).
The colors $V-I$ are similarly randomly shifted with the dispersion
\begin{equation}
\epsilon_{V-I} =\sqrt{\delta(V-I)_H^2+\delta(V-I)^2},
\label{errorcol}
\end{equation}
where $\delta(V-I)_H$ is the uncertainty on the color from the
HIPPARCOS catalog and $\delta(V-I)$ is given by Eq. (\ref{photoerr}).
\item
$\epsilon_{EROS} ({\bf m})$ is the probability to detect
a star with apparent magnitudes ${\bf m}$ in the EROS catalog (see Fig. \ref{effdet}).
Here this probability is a function of $B_{EROS}$ only, which is related
to the absolute magnitudes and to the distance $D$ as follows:
\begin{eqnarray}
&&B_{EROS} = V-0.4(V-I) \label{beros} \\
&=&\mu(D) + M_V+A_V(D)
-0.4(M_V+A_V(D)-M_I -A_I(D)). \nonumber
\end{eqnarray}
\end{itemize}

\subsection{Mass density distributions}
\label{section:mass-distrib}
In this section, we describe two mass distribution
models used to scale the local densities of lenses and sources along the line of sight.
We note the different status of the thick disk: it is considered hypothetical
within the framework of the first model (so-called simple) since it is a pure
hidden matter contribution; on the other hand, it is considered as one of the components within the framework of the
second model (Besan\c{c}on).
\subsubsection{Simple tunable Galactic model}
\label{section:model}
In this model, which is slightly modified (updated) from the so-called model1 we used in \cite{BS7ans},
the mass density of the Galaxy is described with a thin disk and a central
bar structure. The disk is modeled by a double exponential
density in galactocentric cylindrical coordinates
\begin{eqnarray}
\label{densitedisk}
\rho_{D}(r,z) = \frac{\Sigma}{2H} \exp 
\left(\frac{-(r-R_{\odot})}{R} \right) \exp 
\left( \frac{-|z|}{H} \right) \ ,
\end{eqnarray}
where $\Sigma=\mathrm{50 \Msol pc^{-2}}$ is the column density of the disk at
the solar radial position $R_{\odot}=8.3 \mathrm{kpc}$ (\cite{Brunthaler}), $H=0.325 \mathrm{kpc}$ is the height
scale, and $R=3.5 \mathrm{kpc}$ is the radial length scale of the disk.
The position of the Sun with respect to the symmetry plane of the
disk is $z_{\odot}=\mathrm{26\ pc\pm 3\ pc}$ (\cite{Majaess}).
The bar is described in a Cartesian frame positioned at the Galactic
center with the major axis $X$ tilted by $\Phi = 13\degree$ (\cite{Robin})
with respect to the Galactic center-Sun line, {\it i.e.},
\begin{eqnarray}
\rho_{B} &=&  \frac{M_{B}}{6.57 \pi abc} e^{-r^{2}/2} \ , \ 
r^{4} = \left[ \left( \frac{X}{a} \right)^{2} + 
	       \left( \frac{Y}{b} \right)^{2} \right]^{2} + 
	\frac{Z^{4}}{c^{4}}\, ,
\end{eqnarray}
where $M_{B}=1.7\times 10^{10}\mathrm{\Msol}$ is the bar mass,
and $a=1.49 \mathrm{kpc}$, $b=0.58 \mathrm{kpc}$, and $c=0.40 \mathrm{kpc}$ are the scale length factors.\\
There has been some controversy about the bar inclination $\Phi$;
in particular, the EROS collaboration (\cite{Hamadache}) published an erroneously high value
($\Phi = 49\degree \pm 8\degree$) deduced from the variation of the mean
distance to the red giant stars with the Galactic longitude.
This mean distance was confused with the distance to the
bar major axis, but this view is only correct for a zero width bar.
As a consequence, the value of $\Phi$ was strongly overestimated,
since as soon as the bar is elliptic, the barycenters of the stars along
the line of sight do not coincide with the bar main axis (\cite{Lopez}).
Moreover, this difference between the barycenter line and the main axis
increases when  $\Phi$ decreases and when the width of
the bar increases.
Correcting this wrong view, we checked that the EROS red giant clump distance
measurements are in fact compatible with the low values of
$\Phi$ recently published (\cite{Robin}, \cite{Wegg}), as discussed in the
following sections.

The hypothetical thick disk is also considered in our model, and we fit its
fractional contribution $f_{thick}$ to the Galactic structure ($f_{thick}=1$ would correspond to fully
baryonic Galactic hidden matter).
This disk is modeled as the
thin disk (Eq. (\ref{densitedisk})), with $\Sigma_{thick}=\mathrm{35 \Msol pc^{-2}}$, $\mathrm{H_{thick}=1.0 kpc}$, and $\mathrm{R_{thick}=3.5 kpc}$.

The IMF of the stellar population is taken from \cite{Chabrier 2004} (Eq. (\ref{chabrierfunction})).
We already mentioned that we expect the microlensing duration to be
especially sensitive to the low-mass side of the IMF of the lens population.
We therefore define a tunable function for the low-mass side IMF ($m\le \Msol$),
by introducing a parameter $m_0$ (with value $m_0=0.2\Msol$ for the regular Chabrier IMF):
\begin{equation}
\xi(\log m/\Msol)=
0.093\times exp\left[\frac{-(\log m/m_0)^2}{2\times (0.55)^2}\right],\,for\,m\le \Msol
\label{massfunction}
\end{equation}

and we fit this parameter to our microlensing duration data in Section \ref{section:fitting}.

We use the following kinematical parameters:
\begin{itemize}
\item
The radial (axis pointing toward the Galactic center), tangential and perpendicular
solar motions with respect to the disk are taken from (\cite{Brunthaler}),
\begin{equation}
v_{\odot r}=11.1^{+0.69}_{-0.75},\ v_{\odot \theta}=12.24^{+0.47}_{-0.47},\ v_{\odot z}=7.25^{+0.37}_{-0.36}\ \ 
({\rm km}/{\rm s}).
\end{equation}
We found that the microlensing duration distribution obtained in our simulation is
almost insensitive to the exact values of these parameters.
\item
The global rotation of the disk is given as a function of the
galactocentric distance by
\begin{equation}
V_{rot}(r) = V_{rot,\odot} \times \left[ 
1.00767 \left( \frac{r}{R_{\odot}} \right)^{0.0394} + 0.00712 \right] \ ,
\end{equation}
where $r$ is the projected radius (cylindrical coordinates) and
$V_{rot,\odot} = 239\pm 7$ {\rm km/s} (\cite{Brunthaler}).
\item
The peculiar velocity of the (thin or thick) disk stars is described by
an anisotropic Gaussian distribution with the following
radial, tangential, and perpendicular velocity dispersions (\cite{Pasettothin}
and \cite{Pasettothick}):
\begin{align}
\label{velocity}
\sigma_r^{thin} &= 27.4\pm 1.1\ km/s &\sigma_r^{thick} = 56.1\pm 3.8\ km/s \nonumber \\
\sigma_{\theta}^{thin} &= 20.8\pm 1.2\ km/s &\sigma_{\theta}^{thick} = 46.1\pm 6.7\ km/s \\
\sigma_z^{thin} &= 16.3\pm 2.2\ km/s &\sigma_z^{thick} = 35.1\pm 3.4\ km/s. \nonumber
\end{align}
We also found that the microlensing duration distribution is
insensitive to the exact values of these parameters.
\item
The velocity distribution of the bar stars is given by the combination of a global rotation (\cite{Fux}, \cite{rotbar})
\begin{equation}
\Omega_{bar}= 39 km\pm 3.5\ s^{-1}\ kpc^{-1}
\label{barvelocity}
\end{equation}
with a Gaussian isotropic velocity dispersion distribution characterized by $\sigma_{bar} \sim 110 \ {\rm km/s}$.
We found that the mean duration of microlensing events toward $\gsct$, which is the only line of sight crossing the bar, is almost
insensitive to $\Omega_{bar}$, mainly because the global rotation velocity is almost tangent to this line of sight.
\end{itemize}

\subsubsection{Besan\c{c}on Galactic model}
In this model (\cite{besancon}, with updated parameters from \cite{Robin}), the distribution of the matter in the Galaxy is described by the
superposition of eight thin disk structures with different ages, a thick disk component, and a central (old) bar structure
made of two components (\cite{Robin}).
We considered the updated model from (\cite{Robin}) that appears to be specifically adapted to the Galactic plane,
and chose the fitted parameters associated with a two ellipsoid bar (Freundenreich (S) plus exponential (E) shapes).
All the parameters from this model can be found in the Appendix B, to enable any useful comparison with our simple model.

\subsubsection{From the local CMD and mass density to the stellar distribution}
The mass densities are then converted into stellar number densities and
distributed according to our debiased HIPPARCOS-CMD (Section \ref{section:unbias}).
The number density of stars scales with the stellar mass density,
such that the total number density of stars within $0<M_V<8$ equals the total mass density within
the corresponding mass interval $[0.65,2.8]\Msol$, divided by the mean stellar mass in this
interval, as computed from the IMF.
We finally take into account the fact that $\sim$ 2:3 of those stars are in binary systems, as
discussed in Section \ref{section:check}.
This 2:3 poorly known factor and the exact mass to stellar number ratio can both
be absorbed in a global renormalization factor, and
our simulated catalog has been tuned to precisely reproduce the local (debiased) observed HIPPARCOS-CMD.

We have now in hand the full description of stellar number densities according to the mass densities
and the debiased HIPPARCOS-CMD, which is our initial ingredient to simulate EROS-like CMDs.

\subsection{Extinction}
\label{section:absorption}
We now have to consider the absorption model to simulate the effects of distance and reddening
of the sources in expressions (\ref{catcont}) and (\ref{beros}).

\begin{figure}[htbp]
\begin{center}
\includegraphics[width=8.5cm]{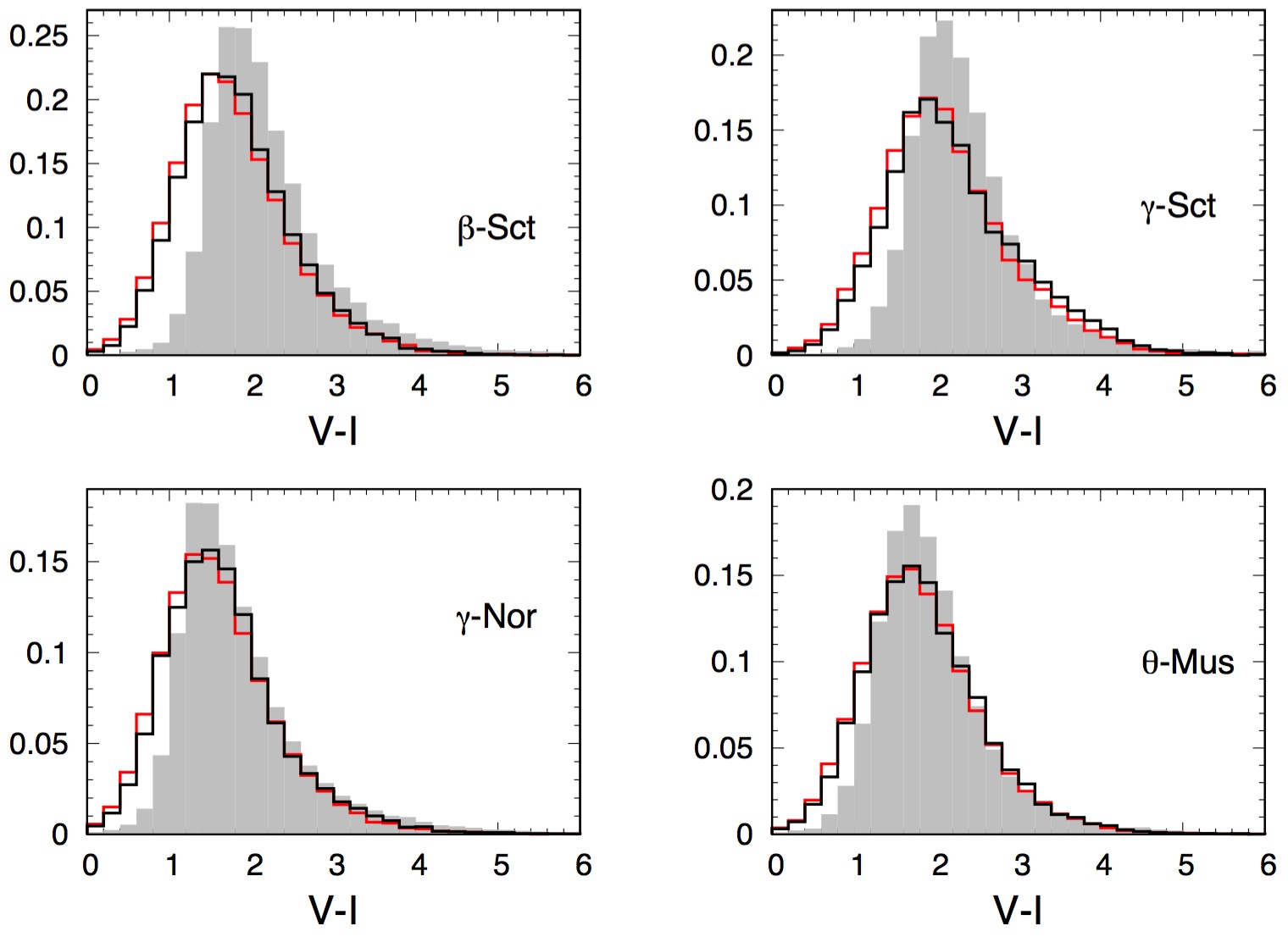}
\caption[]{
The $V-I$ observed (gray histograms) and the simulated distributions (simple model in black, Besan\c{c}on model in red)
for the bright stars (with $I<18.4$),
using the Marshall table without systematic/statistical uncertainties (to be compared with Fig. \ref{HRsim}, bottom).
}
\label{absorption}
\end{center}
\end{figure}

After generating the position and type of a star, we estimate the extinction due to
dust along the line of sight using the table provided by (\cite{Marshalltable}). This
3D table provides $A_K$, the extinction in $K_S$ in the $(b,l)=(\pm 10\degree, \pm 100\degree )$
domain, up to $\sim 15\mathrm{kpc}$, with $0.1\degree $ angular resolution and $0.1\, \mathrm{kpc}$ distance resolution.
We use the following relations to transpose the $A_K$ into $I$ and $V$ passbands
\begin{equation}
A_V=8.55\times A_K,\  A_I=4.70\times A_K, A_{V-I}=3.85\times A_K.
\end{equation}
We compared the extinctions from this table with the 2D table of (\cite{Schlegel}) (through extrapolation
at infinite distance), which is notoriously imprecise toward the Galactic plane,
and with the calculator of (\cite{Schlafly})\footnote{$https://ned.ipac.caltech.edu/help/extinction\_law\_calc.html$}.
We found that up to $\sim 5\,\mathrm{kpc}$, the extinctions in $I$ from \cite{Schlafly} are compatible with the Marshall table, although systematically lower. At larger distances, the estimates depart from each other, and extrapolations
at large distance from Marshall table are much larger than estimates from both \cite{Schlegel} and \cite{Schlafly}.
Nevertheless, as discussed in section \ref{section:syste-ak}, we found it necessary
to correct the extinctions of the Marshall table for systematic
and statistical uncertainties, to get synthetic CMDs of $I<18.4$ stars
that correctly match the observed CMDs (compare Fig. \ref{absorption} with Fig. \ref{HRsim}); indeed, because of the large
multiplicative factor relating $A_V$ and $A_I$ to $A_K$, a small error on $A_K$ has a very significant impact
on the apparent position of a star in our CMD.
Fig. \ref{actorsdist} shows the average extinctions in $V$ along the lines of sights as a function of the distance
to the source, after tuning the model parameters according to our fitting procedure.

\subsubsection{Blending}
\label{section:blending}
We know from the comparison of the EROS images with the HST images (\cite{BS7ans})
that $\sim 60\%$ of the $I<16$ objects and $\sim 70\%$ of the $I>16$ objects
detected by EROS are blends. This blending effect is different than
the binary blend mentioned at the end of Appendix A.
This effect, due to the EROS low separation power, is accounted for
by randomly decreasing the magnitudes of $60\%$ of the faint stars (resp. $70\%$ of the bright)
according to a  Gaussian distribution centered on $-0.07$, with $\sigma=0.25$ (resp. 0.13),
troncated at zero.

In principle, blending also contributes to reduce the number of detected objects with respect
to the predictions based on the HIPPARCOS catalog.
As for the binary blend, this effect can be absorbed in a global renormalization factor.

\section{Comparing the EROS observations with
simulated populations and microlensing expectations}
\label{section:comparison}
Our aim is now to tune and compare the Galactic models with the
observations toward the four Galactic disk lines of sight
(characterized by the corresponding EROS catalogs noted $C$).
We use all the available observables for this purpose as follows:
\begin{itemize}
\item
The four color-magnitude distributions (CMD) of stars brighter than $I=18.4$,
which is the most reliable part of the EROS-CMD. The observable variables we
consider are derived from the projected magnitude and color distributions:
the total stellar densities $\rho_{*}$
\footnote{After noting that the slopes of the magnitude distributions seem universal,
we concluded that the integrated stellar number density $\rho_*$ carries all the information on this distribution.},
and the first moments $\overline{V-I}$ and $\sigma_{V-I}$ of the $V-I$ distribution\footnote{These variables have the
advantage that it does not depend on an arbitrary binning.}.
\item
The measured optical depths $\tau(C)$ (\cite{BS7ans})
toward the four catalogs $C$ (Table \ref{tabfields}).
\item
The measured means $\overline{t_E}(C)$ (\cite{BS7ans}) (Table \ref{tabfields}).
The poor available statistics convinced us not
to use the $\sigma_{t_E}$ parameter in our fitting
procedure, since it is affected by such a large uncertainty that it is essentially not constraining.
\end{itemize}
For quantitative statistical comparisons based on $\chi^2$ studies,
we need good control of the uncertainties on these observables.
The $\tau$ and $\overline{t_E}$ uncertainties
are provided in \cite{BS7ans}.
Table \ref{tabfields} summarizes the numerical data
toward the EROS monitored populations that we
use for the comparison with a simulation (apart $\sigma_{t_E}$).

\subsection{Simulation of the CMDs}
\label{section:syste-ak}
All the relevant information is already given in Sect. \ref{section:generation}.
Here, we briefly summarize the different stages
to simulate EROS CMDs from various models or parameters.

The stellar absolute magnitudes and colors are first randomly chosen according to the
HIPPARCOS unbiased color-magnitude density diagram
of stars with $0<M_V<8$ (Fig. \ref{Hipp_50pc} bottom).
Generated magnitudes are then shifted to take into account
the blending described in Sect. \ref{section:blending}, as well as
the HIPPARCOS parallax uncertainties and EROS photometric
uncertainties (Eqs. (\ref{errorI}) and (\ref{errorcol})).

To estimate the integral in expression (\ref{catcont}),
we generate the distance distributions of stars according to the mass density
distributions of each Galactic structure (bar, thin disk, or thick disk).
The EROS stellar apparent magnitudes and colors are estimated from the
absolute magnitudes, the distances,
and take into account the absorptions tabulated (in $K_S$) at the position randomly chosen
within the EROS fields (see Sect. \ref{section:absorption}).
After this stage, we obtain the apparent color-magnitude distribution of
the stars before detection.
Finally, the contribution of each generated star is weighted
by the EROS stellar detection efficiency $\epsilon_{EROS} ({\bf m})$, which is parametrized
as a function of $B_{EROS}$ (figure \ref{effdet}).

As mentioned in Sect. \ref{section:absorption}, to successfully fit the CMDs
we had to introduce the hypothesis of a systematic uncertainty on $K_S$ changing
with the catalog, $\Delta A_K(C)$, and a random uncertainty with  constant width $\epsilon_{A_K}$,
within the tabulated data. Since the table does not provide uncertainties, we used this hypothesis as the
simpliest way to make our simulation compatible with the observations (A.C. Robin, priv. comm.).
Then the $\Delta A_K(C)$ and $\epsilon_{A_K}$ parameters were tuned
together with the Galactic parameters to
obtain synthetic CMDs that fit the observed CMDs (see below).

\subsection{Simulation of microlensing}
\label{section:simulation}
The previous procedure, based on the synthesis of the color-magnitude diagrams,
allows us to simulate the EROS catalogs of sources.
To simulate the microlensing process for these catalogs,
we also need to synthesize the population of lenses, containing all massive objects
regardless of their visibility. The local lens density population is therefore simulated with
the appropriate IMF (depending on the Galactic structure and on the model)
scaled with the local mass density. The transverse velocity distribution needed to
simulate the microlensing event durations is obtained from the
combination of the velocity distributions from the disk(s) and the
bar, according to their respective local mass contributions.
Finally, we take into account the impact of the time sampling by simulating the microlensing
detection efficiency according to Fig. \ref{efficiency}.

\section{Fitting procedure}
\label{section:fitting}
Our simulation program allows us to produce
the CMDs and microlensing distributions toward our 4 catalogs labeled (C),
with any choice of Galactic parameters.
We detail below the procedure developed for our simple tunable model, which we also
used to probe the Besan\c{c}on model (with no tuned parameter other
than the systematic uncertainties of the interstellar absorptions).
\subsection{Fit and tuning of the simple model}
We examined the following 16 observables (4 per target C)
$\rho_*(C)$, $\overline{V-I}(C)$, $\tau(C)$, and $\overline{t_E}(C)$
as a function of the following parameters, around their nominal values:
$\epsilon_{A_K}$, the random uncertainty on the extinctions $A_{K}$ provided by the table
from \cite{Marshall} for each generated stellar position;
$\Delta A_K(C)$, the systematic uncertainty on $A_{K}(C)$, depending on the catalog$(C)$;
the Galactic bar inclination $\Phi$ (nominal value $\Phi=13\degree$);
and the (hypothetic) thick disk contribution, which is parametrized by the fraction $f_{thick}$ of the
thick disk considered in (\cite{BS7ans}). This contribution is modeled like the thin
disk (see Eq. (\ref{densitedisk})), with
$\Sigma_{thick}= f_{thick} \times 35\, \mathrm{\Msol pc^{-2}}$, $H_{thick}=\mathrm{1.0\, kpc}$, $R_{thick}=\mathrm{3.5\, kpc}$,
and velocity dispersions given by Eq. (\ref{velocity}).

To benefit from the exclusive time information $\overline{t_E}(C)$ provided by the microlensing data,
we also considered some specific parameters that are expected to impact
the microlensing optical durations. 
First, the low-mass part of the IMF, which we generalized from \cite{Chabrier 2004}
through parameter $m_0$ (nominal value $m_0=0.2$) (Sect. \ref{section:model}).
Second, we explored the sensitivity to the peculiar velocities of the
microlensing actors through a scaling of the velocity dispersions
reported in expression (\ref{velocity}). We found that our simulation is insensitive to
such a scaling, therefore confirming that orbital velocities dominate the relative transverse motions.
Third, for completeness, we also tested the sensitivity of $\overline{t_E}$ with the global rotation of the bar (Eq. \ref{barvelocity}) and
found almost no sensitivity; this is mainly because the bar rotation is almost tangent to the
line of sight of $\gsct$, which is the only line of sight that crosses the bar structure.

\subsubsection{Sensitivity of the observables with respect to the Galactic parameters}
We used our simulation to establish the sensitivity of the observables with
the variations of the different parameters, and
we made the following observations.

We find that only the simulated observables from the low longitude fields ($\beta$ Sct and $\gamma$ Sct)
are sensitive to the variations of $\Phi$, when we test for very large changes,
but they are insensitive to few degree variations around the nominal value $\Phi=13\degree$.
As a consequence, we exclude $\Phi$ from our fit.

At first order, the absorption random shift dispersion $\epsilon_{A_K}$, with respect to the tabulated values,
is assumed to be the same for all fields, and the widths of the
four color distributions $\sigma_{V-I}(C)$ are found to be disconnected from the other
observables and parameters. We therefore directly fit $\epsilon_{A_K}$ by minimizing the differences between
$(\sigma_{V-I}^{obs.}(C))^2$ and the width combination $(\sigma_{V-I}^{sim.(0)}(C))^2 + (3.85*\epsilon_{A_K})^2$,
where the 3.85 factor comes from the relation $A_{V-I}=3.85\times A_K$ (see Sect. \ref{section:absorption}),
and the $\sigma_{V-I}^{sim.(0)}(C)$ values are obtained with a simulation that assumes $\epsilon_{A_K}=0$.
The value that minimizes the sum on $(C)$ is
$\epsilon_{A_K}=0.085$, which we assume to be independent of the catalog $C$. We use
this value in the subsequent simulations.

The Chabrier-like IMF parameter $m_0$
and the observables $\overline{t_E}(C)$ are also disconnected from the other observables and parameters.
We therefore make a separate (sub-)fit for these parameters, by minimizing
\begin{equation}
\chi^2_{t_E} = \sum^{catalogs}_{C} \frac{(\overline{t_E}^{sim}(C)-\overline{t_E}^{obs}(C))^2}{\sigma^2_{t_E}(C)}
\end{equation}
with respect to $m_0$,
where the suffixes $sim$ and $obs$ refer to the simulated and observed catalogs.

The observables $\rho_{*}(C)$,
$\overline{V-I}(C)$, and the
microlensing optical depths $\tau(C)$ (12 observables) depend only on $f_{thick}$ and on the systematics
$\Delta A_K(C)$ (5 parameters). We performed a combined fit by minimizing the sum of
$\chi^2_{\rho_*}$, $\chi^2_{\overline{V-I}}$ and $\chi^2_{\tau}$, which is defined similar to $\chi^2_{t_E}$,
but since we have to take into account common systematics, some of the covariant matrices are not diagonal.

In our minimization procedure, we used the first order developement of the observables as functions
of the parameters to be fitted, from the derivatives computed with our simulation.
This allowed us to perform the fit with acceptable computing time, considering the very long runs needed
for each model configuration.
\subsubsection{Systematic and statistical uncertainties}
We have carefully established the budget error for each observable as follows.

For the $\rho_{*}(C)$ budget error, we have to take into account the uncertainty of $\sim 5.3\%$ on the size of the
effective EROS field
and the consequences of the $0.07$ magnitude EROS calibration uncertainty.
The impact of this calibration uncertainty on $\rho_{*}(C)$ has been estimated from
the published EROS-CMD tables, by changing
the position of the $I<18.4$ magnitude cut by the $0.07$ systematics.
We found that the uncertainty on $\rho_{*}(C)$ due to this calibration error is $\sim 5\%$.
The final systematics results from the quadratic addition of both uncertainties
($7.3\%$) and since it is a multiplicative systematics, it has to be considered as an uncertainty on
a global normalization
$\alpha$; we therefore use a standard procedure to include the extra parameter $\alpha$ and fit the product
$\alpha\times\rho_{*}^{sim}$ with $\rho_{*}^{obs}$.
We adopt $15\%$ as the statistical uncertainty on $\rho_{*}(C)$, which is dominated by residual uncertainties
from the absorption model and blending effects.

For $\overline{V-I}(C)$ , we have to account for the systematics due to calibration
uncertainties on both $R_{EROS}$ and $B_{EROS}$, thus giving a global systematics of
$0.16\, \mathrm{mag.}$. In the covariance matrix associated with the fit minimization,
this additive systematics, which is common to the four directions, contributes as a full matrix, to be added to the usual diagonal
matrix built from the residual statistical uncertainty that is estimated to be $0.15\, \mathrm{mag.}$.

Statistical uncertainties from the EROS-CMD Poissonian fluctuation propagation are estimated
as explained in the header of the published EROS-CMD (\cite{BSweb}).
Considering the large statistics available in the EROS database,
we can neglect the uncertainties due to the Poissonian fluctuations of the number of stars in the original EROS
histogram used to produce the CMDs.

As a conclusion, the uncertainties on $\rho_{*}(C)$, $\overline{V-I}(C)$ and $\sigma_{V-I}(C)$ are dominated by the impact
of the calibration uncertainties and the residual uncertainties from blending and absorption effects
discussed above. The values used for the fit are summarized in Table \ref{results}.

\subsubsection{Results from the fit}
We remind that the fit is done with the best value for the random uncertainty on the tabulated absorptions $A_{K}$: $\epsilon_{A_K}=0.085$.
The best fit is obtained with the following parameters.

First, regarding absorption systematics, we find
$\Delta A_K(\beta Sct)=0.09\, \mathrm{mag}$,
$\Delta A_K(\gamma Sct)=0.04\, \mathrm{mag}$,
$\Delta A_K(\gamma Nor)=0.11\, \mathrm{mag}$,
and $\Delta A_K(\theta Mus)=-0.01\, \mathrm{mag}$.

Second, regarding the fraction of the thick disk, we find $f_{thick}=0.05 \pm 0.6$. This result does not differ from zero, showing
that there is no need for an additional baryonic contribution to the thin disk within the framework of our
simple model.
We also tested the option of a non-luminous thick disk (made of compact unseen objects),
assuming no contribution to the CMD (therefore only impacting the optical depths);
we found $f_{invisible\, thick}=0.5 \pm 0.9$, which is again not significantly different than zero.
From this estimate, we can conclude that the total mass of an invisible thick disk
is smaller than $7\times 10^{10}\Msol$ at $95\% CL$.

\begin{figure}[htbp]
\begin{center}
\includegraphics[width=9.cm]{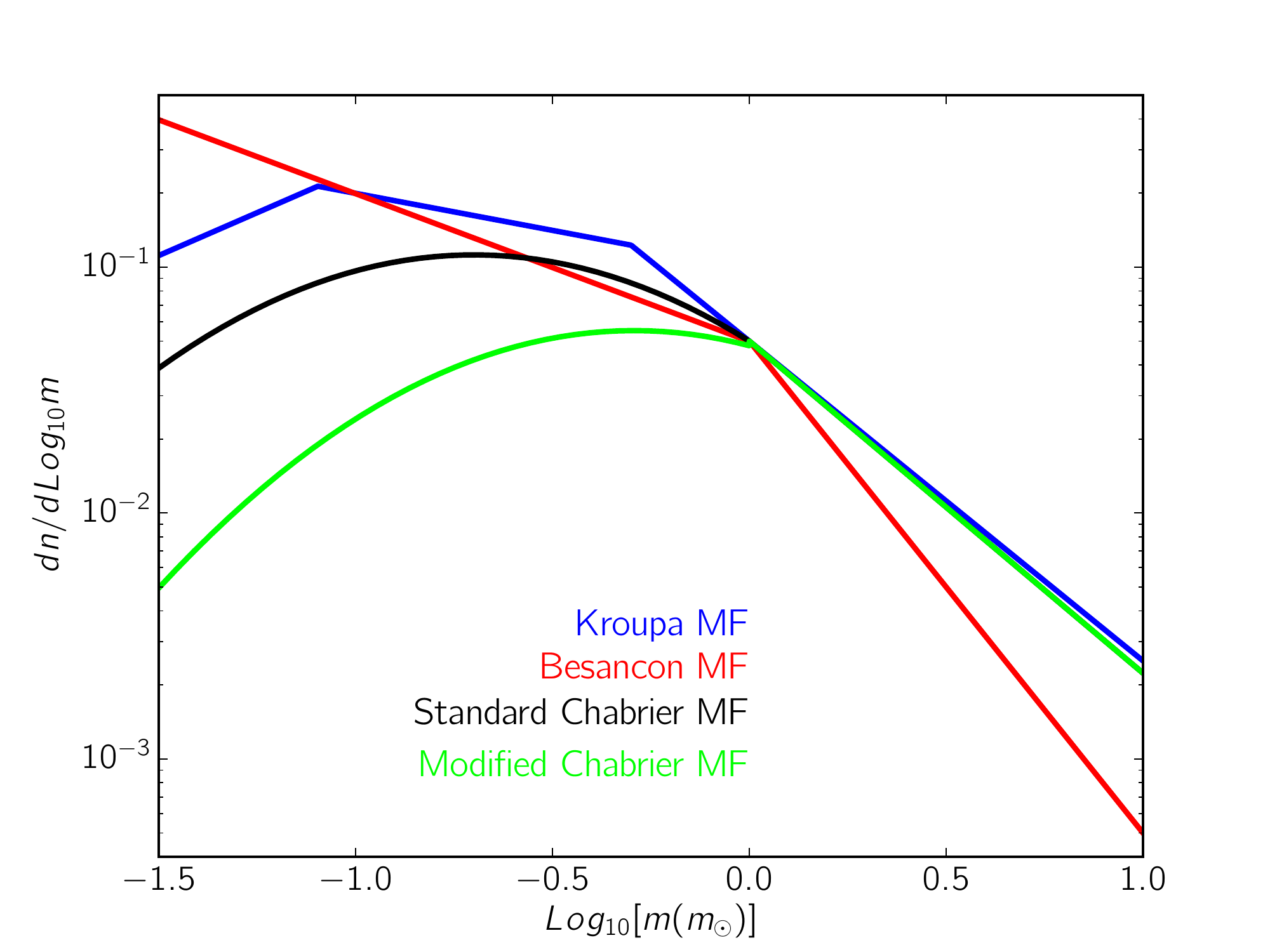}
\caption[]{Different mass functions considered in this paper:
Standard Chabrier (black) corresponds to the local regular Chabrier IMF (Eq. (\ref{chabrierfunction}) with $m_0=0.2\Msol$);
the modified Chabrier ($m_0=0.51\Msol$, in green) gives the best fit for the lens IMF from our simple model.
}
\label{fitdyn}
\end{center}
\end{figure}

Third, regarding the IMF, we find $m_0=0.51\pm 0.25 \Msol$, which is somewhat significantly
different than the $0.2$ nominal value of the local Chabrier IMF (\cite{Chabrier 2004}).
Our observations are therefore
significantly sensitive to the low-mass side of the lens IMF.
This sensitivity belongs to a non-local IMF, since it
concerns only the lenses and not the solar neighborhood.
Fig. \ref{fitdyn} shows both IMFs (the local and best fitted lens-IMF).

For this global fit of the CMDs, optical depths and microlensing durations, we find $\chi^2=6.5$ for 10 degrees of
freedom with a fair repartition between the different types of observables ($\rho_{*}$, $\overline{V-I}$, $\tau$ and $t_E$).

We exchanged in our simple model the Chabrier IMF for the Kroupa IMF (\cite{Kroupa}).
The only consequence to this exchange was a significant decrease in the $\overline{t_E}$ values, as expected from
the larger contribution of low-mass objects (see Fig. \ref{fitdyn} and Table \ref{results}).
This degrades the fit by 7.4 units, showing that the Kroupa IMF is strongly disfavored by our data.

It is clear that a larger statistics of
microlensing events toward the spiral arms would have the capability to better constrain
the thick disk component and the lens-IMF.

\begin{figure}[htbp]
\begin{center}
\includegraphics[width=8.cm]{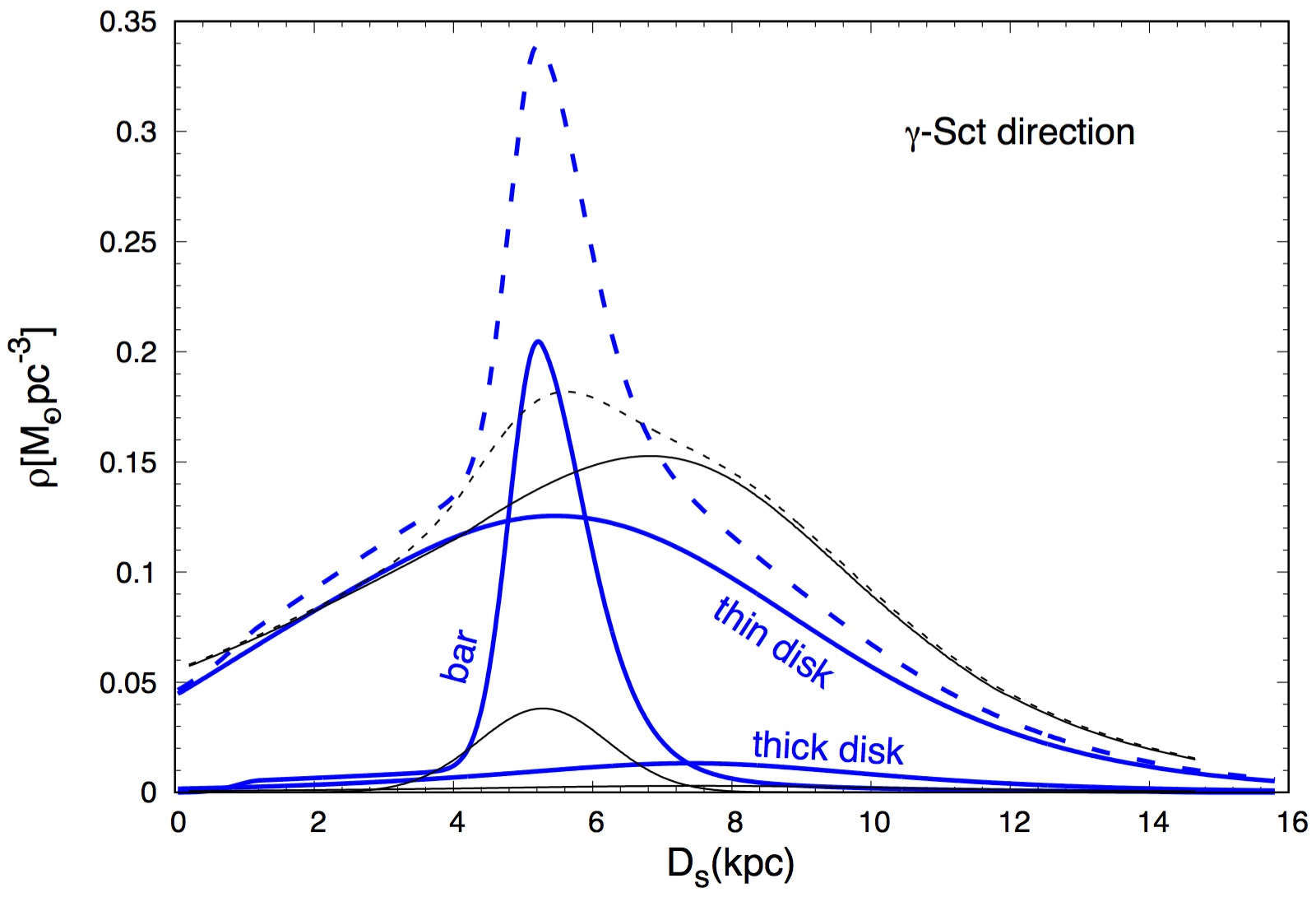}
\caption[]{
Mass-density along the line of sight of $\gamma Sct$ from the various Galactic structures (disks and bar),
as a function of the distance from the Sun for our nominal simple model (thin black lines)  and the
Besan\c{c}on model (thick blue lines). The total densities are shown with dashed lines.
}
\label{densitevsdist}
\end{center}
\end{figure}

Table \ref{results} summarizes the best fit results for our simple model compared with previous simulations
(model 1) considered in \cite{BS7ans}, differing mainly through the extinction description.
\setlength{\tabcolsep}{3pt}
\begin{table}[h!] 
\begin{center}
\caption{ Best fit results on the observables toward the 4 regions monitored 
in the EROS spiral arms program, compared with previous
simulations (model 1) and observations published in \cite{BS7ans}:
Surface density (per square degree) of stars brighter than $I=18.4$, mean and width of CMD color distribution,
number of microlensing events, optical depth, and mean duration.
} 
\label{results}
{\scriptsize
\begin{tabular}{c c c c c c}\hline
 & Target & $\tmus$ &  $\gnor$ & $\gsct$ & $\bsct$  \\
\hline
 & measured  & $0.25\pm .037$ & $0.23\pm .035$ & $0.28\pm .042$ & $0.34\pm .051$ \\ 
$\rho_{*}^{I<18.4} \times 10^6$ & & \multicolumn{4}{c}{$\pm 7.3\%$ common systematics} \\
 & simple model & $0.22$ & $0.26$ & $0.28 $ & $0.32$ \\ 
 & Besan\c{c}on & $0.23$ & $0.26$ & $0.30 $ & $0.33$ \\ 
\hline
 & measured    &$1.95\pm .15$ & $1.86\pm .15$ & $2.36\pm .15$ & $2.20\pm .15$ \\ 
$\overline{V-I}$ & & \multicolumn{4}{c}{$\pm 0.16$ common systematics} \\
 & simple model & $1.83$ & $2.02$ & $2.35$ & $2.13$ \\ 
 & Besan\c{c}on & $1.94$ & $2.11$ & $2.52$ & $2.22$ \\ 
\hline
 & measured    &$0.71$ & $0.78$ & $0.71$ & $0.75$ \\ 
$\sigma_{V-I}$ & simple model & $0.72$ & $0.73$ & $0.83$ & $0.74$ \\
 & Besan\c{c}on & $0.73$ & $0.74$ & $0.81$ & $0.73$ \\
\hline
$N_{event}(u_0<.7)$ & observed  & 3 & 10 & 6 & 3 \\
\\
 & model 1 & 2.8 & 9.9 & 7.1 & 6.3 \\
$\overline{N}_{event}(u_0<.7)$ & simple model & 4.0 & 8.6 & 3.6 & 2.2 \\ 
 & Besan\c{c}on & 4.0 & 9.9 & 3.5 & 2.4 \\ 
\hline
 & measured    &$.67^{+.63}_{-.52}$&$.49^{+.21}_{-.18}$&$.72^{+.41}_{-.28}$&$.30^{+.23}_{-.20}$ \\  
 \\
$\tau\times 10^6$ & model 1 & $0.42$ & $0.52$ & $0.71$ & $0.57$ \\ 
 & simple model & $0.23$ & $0.38$ & $0.43$ & $0.45$ \\ 
 & Besan\c{c}on & $0.22$ & $0.34$ & $0.44$ & $0.40$ \\ 
 \hline
 & measured & $97\pm 75$ & $57\pm 10$ & $47\pm 6$ & $59\pm 9$ \\
 \\
 & model 1 & $73.8$ & $67.9$ & $37.9$ & $60.2$ \\
 $\overline{t_E}$ (days) & simple model & $79.4$ & $54.4$ & $49.1$ & $53.8$ \\
 & with Kroupa IMF & $64$ & $43$ & $38$ & $42$ \\
 & Besan\c{c}on & $68.5$ & $51.9$ & $43.0$ & $49.3$ \\

\hline
\end{tabular}
}
\end{center}
\end{table}
Fig. \ref{densitevsdist} shows the mass density along the line of sight of $\gamma Sct$ resulting from
our simple fitted model.

\subsection{Besan\c{c}on model: tuning the extinctions}
In this section, our purpose is to test the agreement of the Besan\c{c}on model with
the EROS microlensing results.
We used almost the same procedure as above, but fitting only the
uncertainties on the $K$ extinctions.
The best fit is obtained for 
$\epsilon_{A_K}=0.10$,
$\Delta A_K(\beta Sct)=0.14\ mag$,
$\Delta A_K(\gamma Sct)=0.13\ mag$,
$\Delta A_K(\gamma Nor)=0.15\ mag$, and
$\Delta A_K(\theta Mus)=0.04\ mag$.
The global fit has a $\chi^2=8.2$ for 12 d.o.f, with specific contributions of
$\chi^2_{\rho_*}=1.2$, $\chi^2_{\overline{V-I}}=2.2$, $\chi^2_{\tau}=2.8$, and $\chi^2_{t_E}=2.0$.

Not surprisingly, the
values of $\chi^2_{\tau}=2.8$ and $\chi^2_{t_E}$ are worse than those of our simple model, since no parameters are fitted
for the thick disk and the IMF, but the fit is globally satisfying (see Table \ref{results} for the summary of the
fitted parameters and observables).
Fig. \ref{densitevsdist} shows the mass density along the line of sight of $\gamma Sct$ from the Galactic
structures of the Besan\c{c}on model (in blue), resulting from the best fitted extinction.

As for the previous simple model, we also tested the hypothesis of an invisible extra contribution to the thick
disk for this model;
we find that the best fitted value for such a thick disk favors an added contribution of $2.5\pm 4.7$ times the
modeled thick disk ($\chi^2=8.0\, per\, 11\, d.o.f$). Again, there is no significant indication of the need
for such an invisible contribution and the upper limit of a Besan\c{c}on-like thick disk (somewhat thinner
than in our simple model) is $\sim 5\times 10^{10}\Msol$ at $95\% CL$.

\section{Discussion}
\label{section:discussion}
As a preliminary to the discussion, we recall here some of the hypotheses used throughout this paper:
First, we assume the disk to have the same CMD as around the sun; then we rely on the
extrapolation of the extinction map obtained in $K$ band to  $I$ and $V$ bands, and assume
reasonable systematic uncertainties on this map.
\subsection{Comparison with previous results and robustness}
\begin{figure}[htbp]
\begin{center}
\includegraphics[width=8.cm]{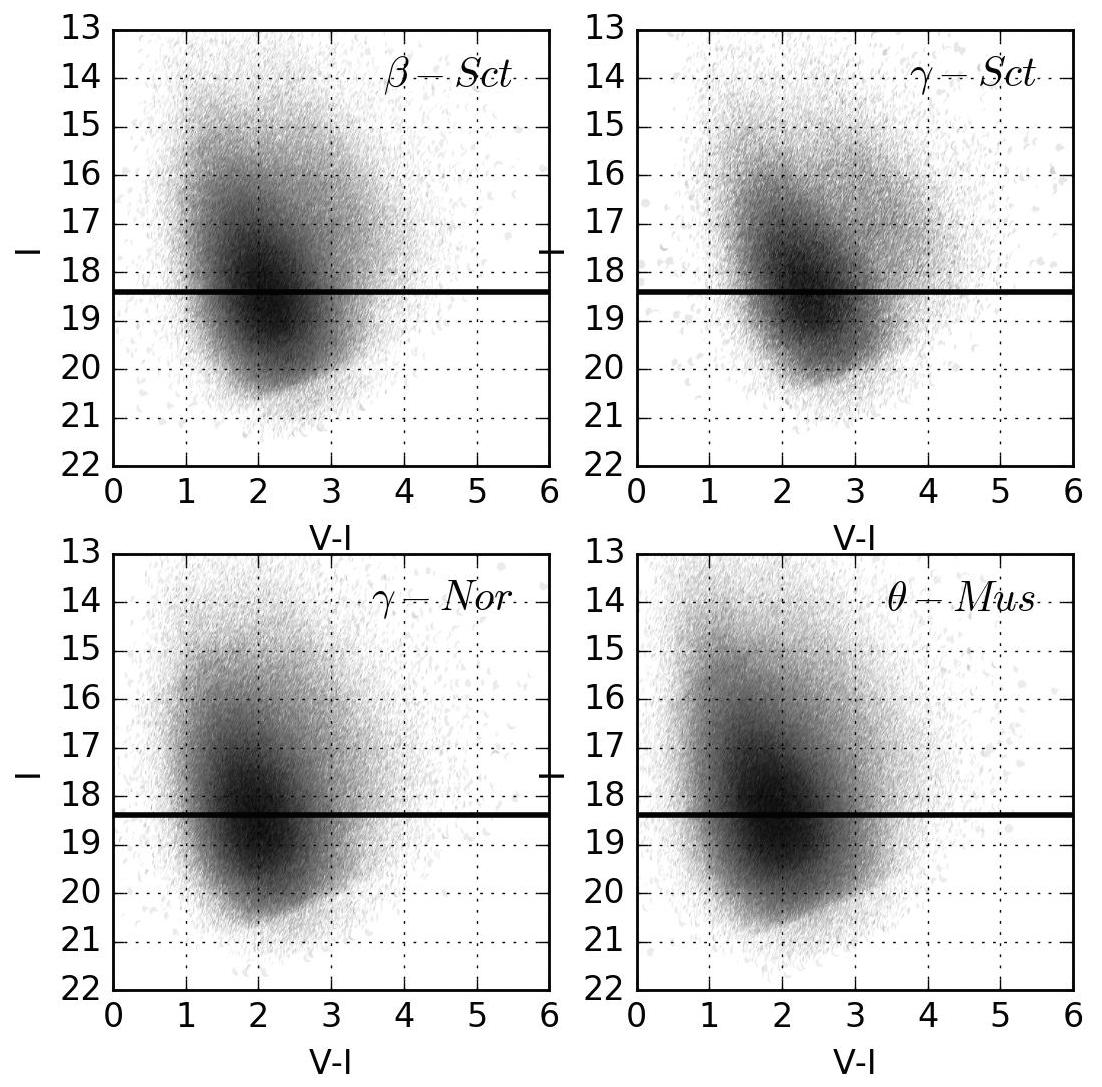} 
\includegraphics[width=8.cm]{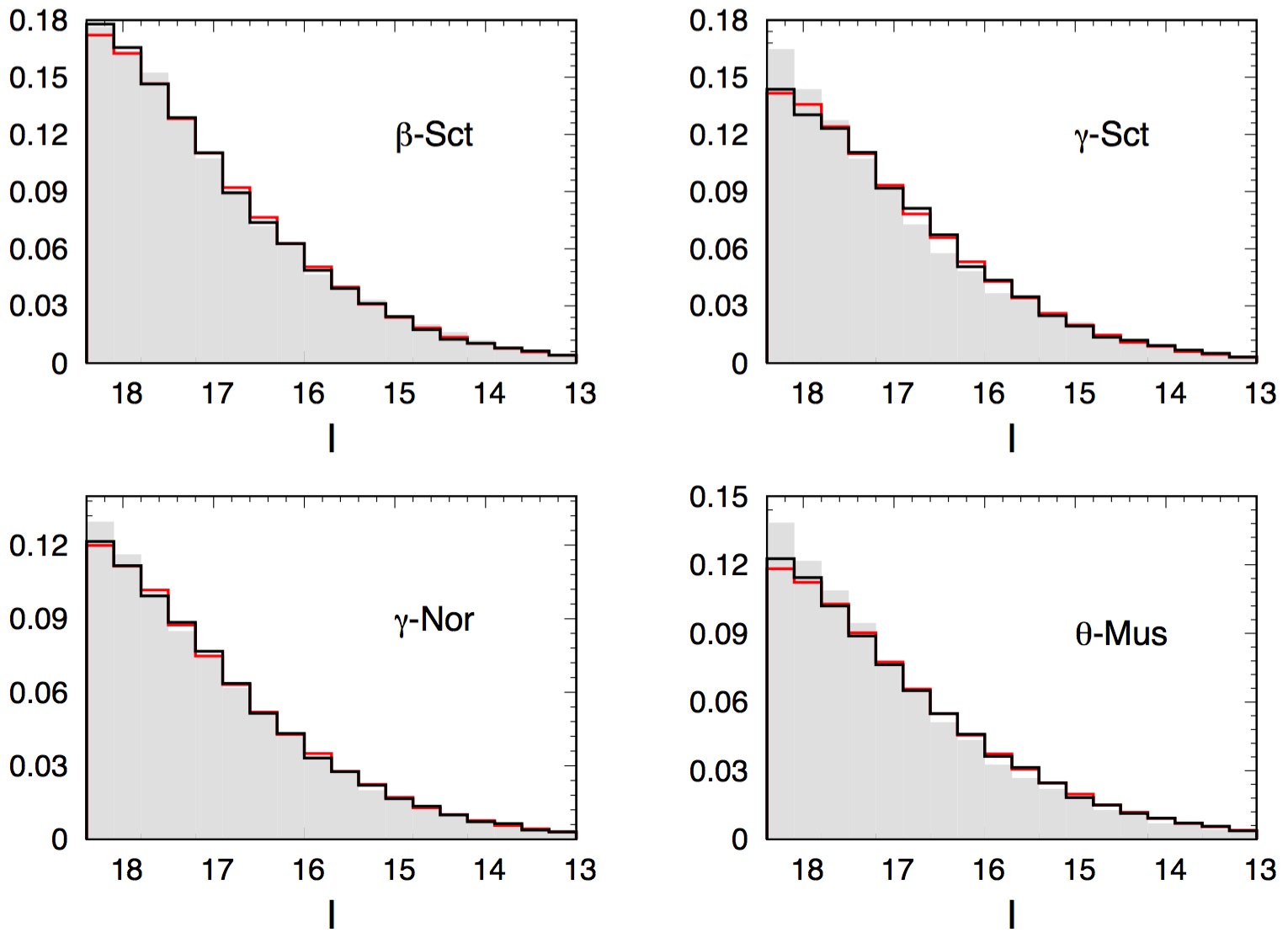}
\includegraphics[width=8.cm]{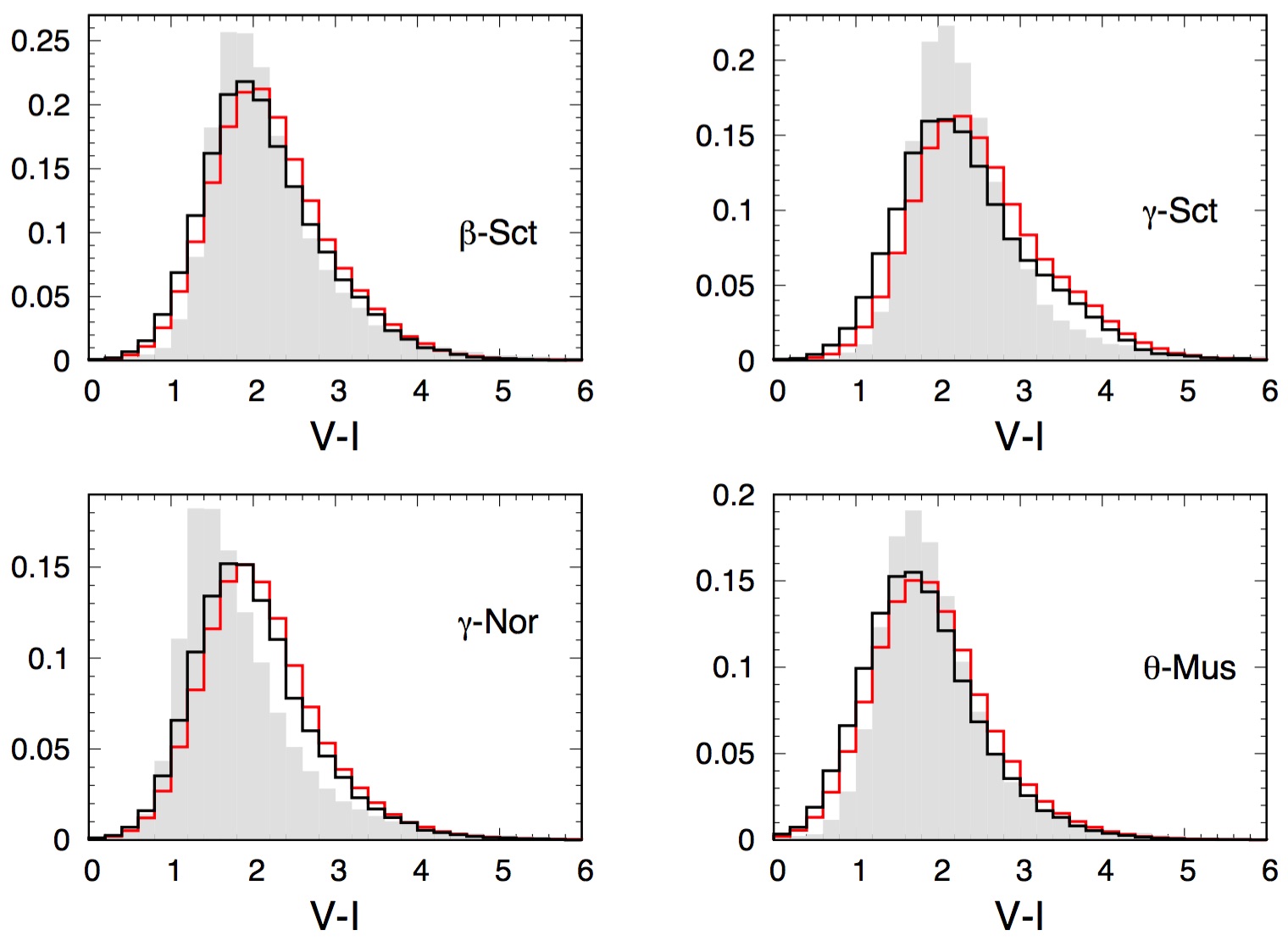}
\caption[]{
Simulated CMDs toward the 4 monitored directions (top)
with the magnitude (middle) and color (bottom) projections for the
stars brighter than $I=18.4$, expressed
in million of stars per square degree per magnitude.
Results from our simple model are plotted with black lines and results
from the Besan\c{c}on model with red lines;
the distributions of the EROS observed populations of bright stars ($I<18.4$)
are superimposed on the projections as light gray histograms.
}
\label{HRsim}
\end{center}
\end{figure}

Figure \ref{HRsim} (to be compared with Fig. \ref{HRdiagrams})
shows that our best fitted models are able to reproduce satisfactorily the observed
CMDs of the ($I<18.4$) stars.
Table \ref{results} shows that the model we used previously (model 1)
was also satisfactory.
We tested the robustness of our results by
changing some of the uncertainties
(systematics and statistics) with unsignificant variations of the best fitted numbers.

Our model now incorporates enough details to allow one to use
the CMD as an observable to be fitted. As a consequence, the main impact of
this type of study, apart from constraining the parameters $f_{thick}$ (for our simple model) and $m_0$,
is to extract information on the underlying stellar populations of sources and lenses.

\subsection{Lens and source populations}
\begin{figure}[htbp]
\begin{center}
\includegraphics[width=8.cm]{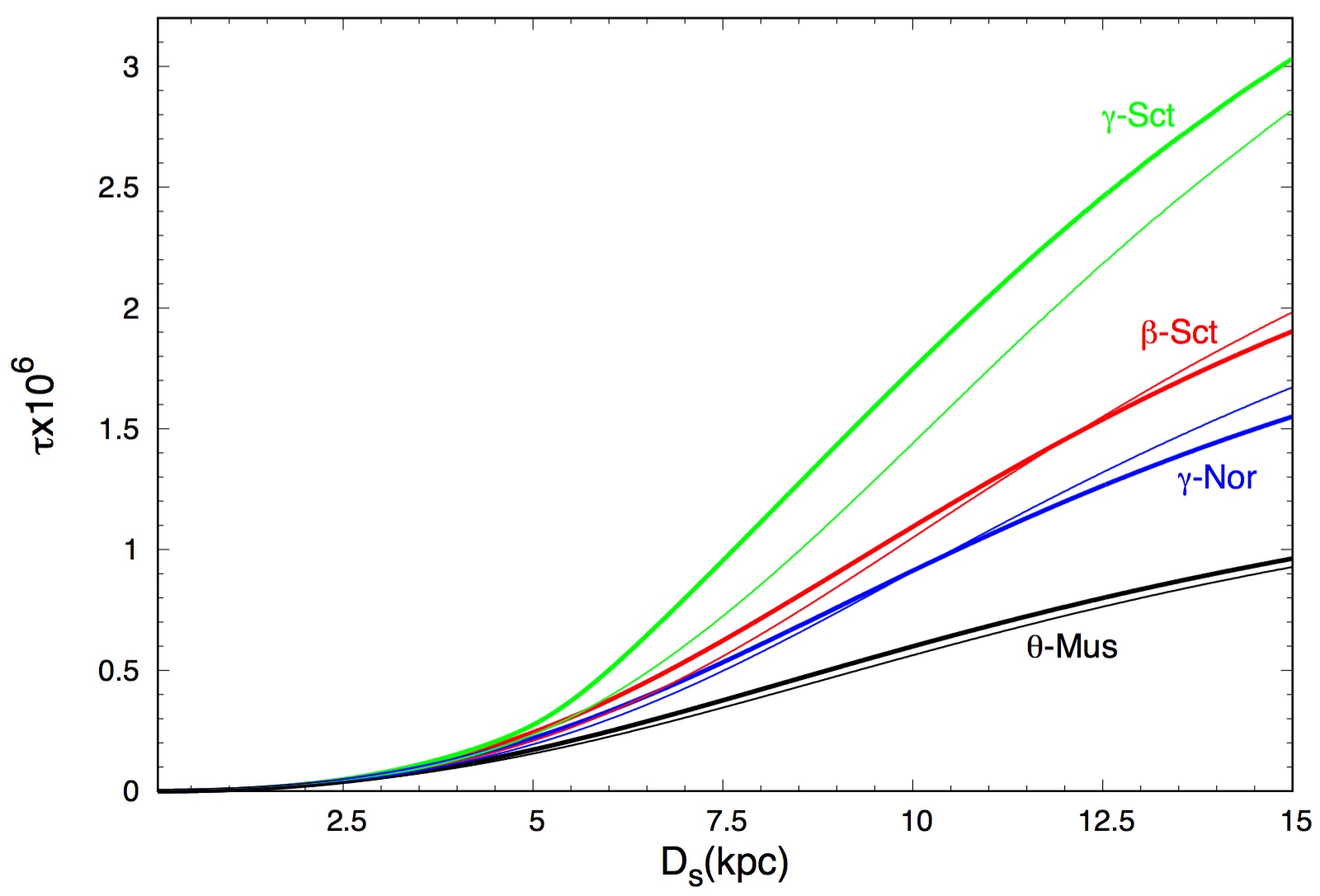}
\caption[]{
Simulated optical depths toward the 4 monitored directions, as a function of the
source distance, for our nominal simple model (thin lines)  and the Besan\c{c}on model (thick lines).
}
\label{optvsdist}
\end{center}
\end{figure}

\begin{figure}[htbp]
\begin{center}
\includegraphics[width=8.cm]{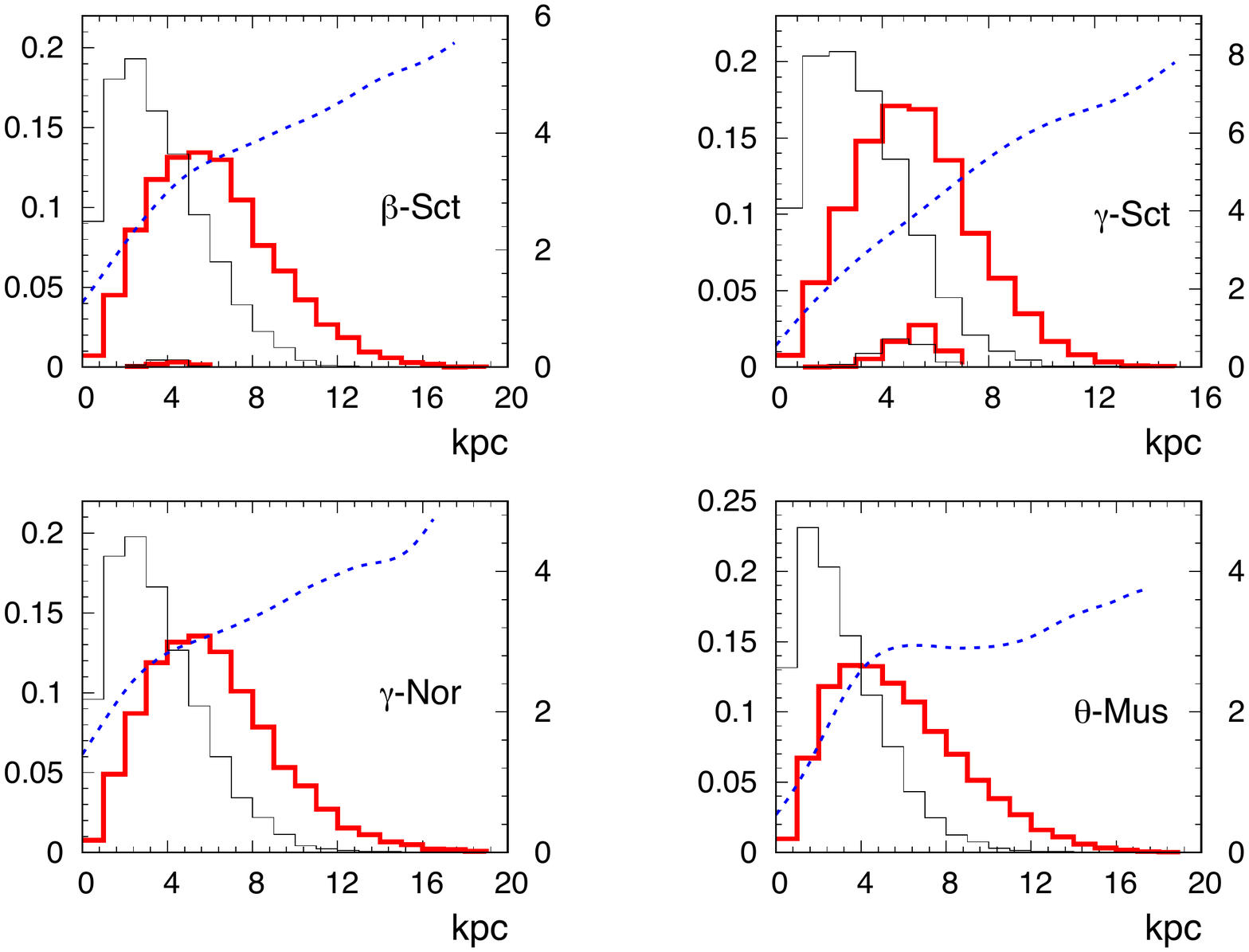}
\includegraphics[width=8.cm]{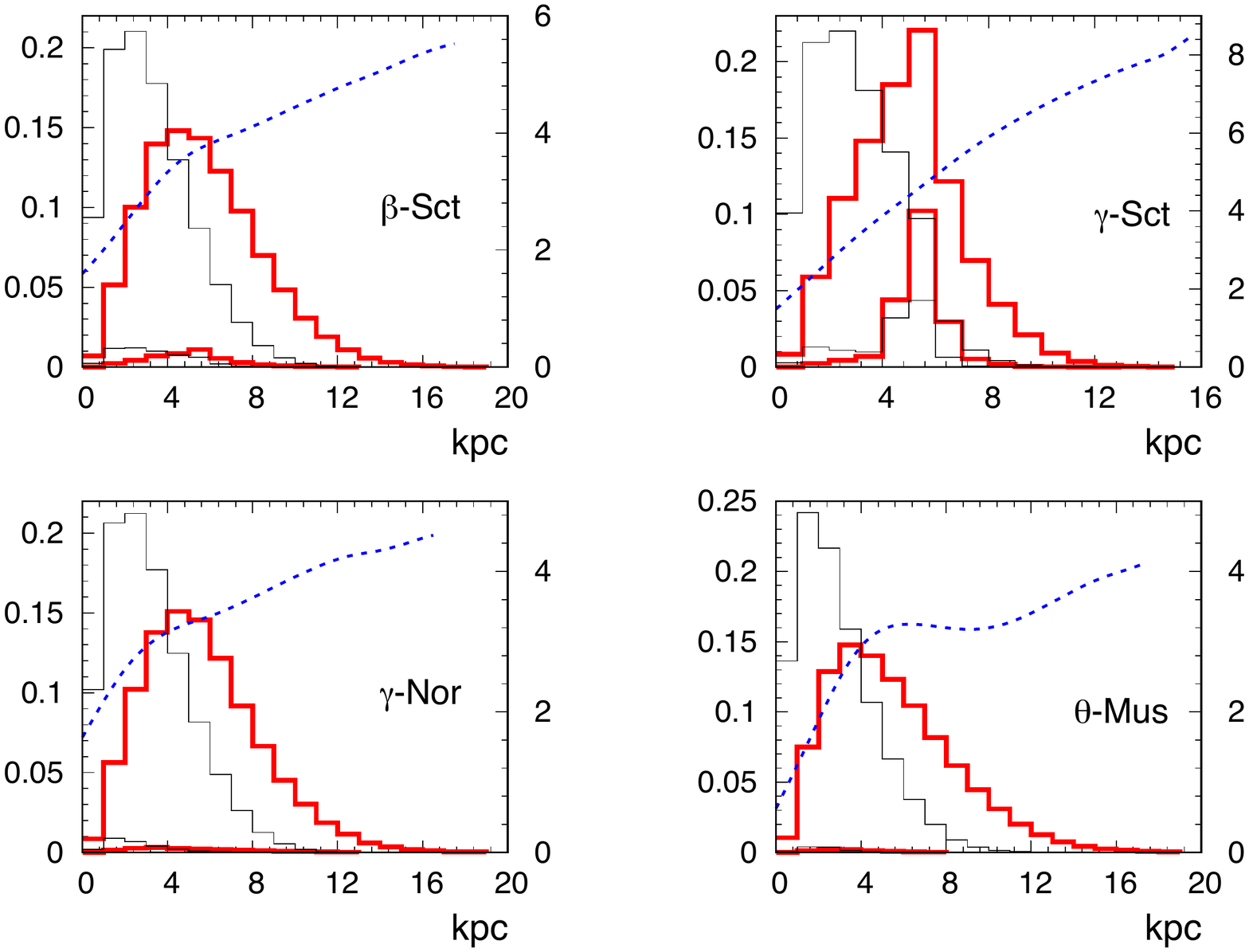}
\caption[]{
Expected normalized distributions of the distances
for the lensed sources ---when taking into account the EROS microlensing detection efficiencies--- (thin lines) and
of the lenses (thick lines) from the simulation of our simple model (upper) and
the Besan\c{c}on model (lower).
The sparsely-populated distributions around $4\,\mathrm{kpc}$ (for $\bsct$ and $\gsct$) correspond
to the contribution of the bar objects.
The dashed curves show, as a function of the distance, the average extinctions
of the stars in the simulated EROS-like catalog (in V magnitude, on the right scale).
It is strongly biased in favor of small extinctions
mainly due to the magnitude selection $I<18.4$.
}
\label{actorsdist}
\end{center}
\end{figure}

Figure \ref{optvsdist} shows the fast variation of the simulated
optical depth along the line of
sight with the distance for the four studied directions and for both models
considered in this paper.
This fast variation of the optical depth with the distance shows that the notion of
catalog optical depth is crucial when dealing with sources distributed along a line of sight.
This notion is not relevant when considering well-defined distance
targets such as LMC, SMC, and M31;  when considering only bright sources toward the Galactic Center,
it is estimated that the relative uncertainty on the bright source’s positions is less than $10\%$ (\cite{distCG})
and it is still possible to ignore the spread of the sources and to use the classical concept of optical
depth up to a given distance for the whole catalog. Previous studies concerning the Galactic spiral
arms (\cite{BS2ans}, \cite{BS3ans}) performed a simplified analysis, by assuming all sources to be at $7Kpc$
to compare the observed optical depth with simple models,
but \cite{BS7ans} started to draw attention to the impact of the source distance spread.
Now it is clear that precise studies in the Galactic plane are needed
to know the distance distribution of the monitored catalog.
Figure \ref{actorsdist} shows the expected distance distributions of the lenses
and sources in the EROS microlensing events 
obtained from our simulation (taking into account the EROS efficiencies).
Again, the source distance distribution illustrates the relevance of the concept of
optical depth toward a population in contrast with the
optical depth up to a given distance.
\subsection{Constraining the Galactic model: The specific contribution of microlensing data}
The good agreement of our Galactic models with
the data shows that there is no need for other or more ingredients.
The Besan\c{c}on model predicts relatively small optical depths, and this
observation is in agreement with the deficit of optical depth
toward the inner bulge directions noticed by MOA-II (\cite{Awiphan}), even if this is not very
significant from our reduced statistics.

We also used our simulation to measure
the domain of Galactic parameters that is compatible with our observations.
We focused on parameters that are expected to impact
the microlensing optical depths or durations, {\it i.e.},
the bar inclination $\Phi$ (nominal value $\Phi=13\degree$);
the thick disk contribution, parametrized by the fraction $f_{thick}$, either visible (for the
simple model) or invisible (for both models);
the disk kinematics for which we explored our sensitivity 
through the scaling of the velocity dispersions (in expression (\ref{velocity}));
and the IMF parameter $m_0$, as defined in Sect. \ref{section:fitting}.

The impact of the Galactic bar is illustrated in Fig. \ref{actorsdist}, where it is clearly visible
that it mainly intercepts the $\gsct$ line of sight;
in the present case, owing to low statistics, our data can only distinguish between a small or a large bar angle,
but cannot refine its current estimate.
Nevertheless, it is clear that systematic microlensing study at relatively small Galactic longitude
is a promising technique to precisely measure the bar inclination.

We show that there is no significant need for an extra thick disk component (visible or invisible);
otherwise, from our data alone there are not enough constraints to exclude its existence and only a $95\% CL$
upper limit on its total mass could be inferred ($\sim 7\times 10^{10}\,\mathrm{\Msol}$ for the simple-model thick disk,
$\sim 5\times 10^{10}\,\mathrm{\Msol}$ for an extra invisible component in the Besan\c{c}on-model thick disk).

We found that we cannot constrain the velocity dispersion ellipsoids of the microlensing actors,
since the transverse velocities involved in the microlensing durations are dominated by the
orbital velocities.

\begin{figure}[htbp]
\begin{center}
\includegraphics[width=9.cm]{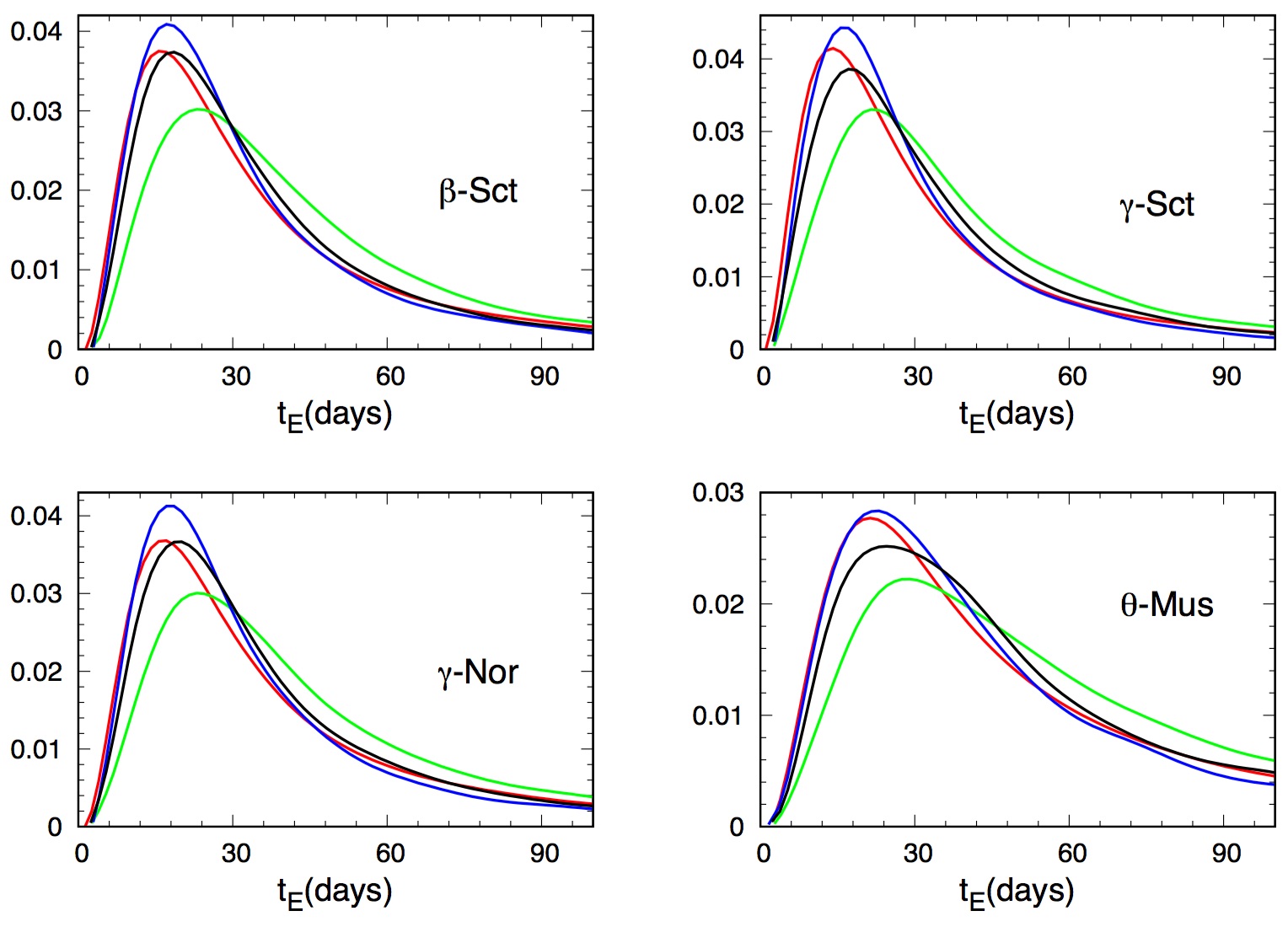}
\caption[]{
Einstein duration $t_E$ distribution of the microlensing events expected by assuming
4 different IMFs: the standard Chabrier (black), the Besan\c{c}on model (red), the modified
Chabrier (with $m_0=0.57$, green), and the Kroupa IMF (blue).
}
\label{distduration}
\end{center}
\end{figure}

Interestingly, we show that microlensing durations can constrain the low-mass end of the mass function
(see Fig. \ref{distduration}), and more importantly, it can provide such constraints for non-local
stellar populations (the disk lens population); this is in contrast with the other techniques, which
can only measure the mass function around the Sun.

The best fitted value we obtain for our parametrized Chabrier-type IMF of the lens population of the disk is
$m_0=0.51\pm 0.25$, which is in relative disagreement (by one standard deviation) with the parameter of
the local mass function ($m_0=0.2$) of the Chabrier model.
This discrepancy originates in the longer mean durations of the observed events
compared with the simulation based on the local IMF. The microlensing technique seems to be
significantly sensitive to the IMF low-mass end.

We find that the Kroupa IMF does not correctly reproduce the mean durations of our microlensing events,
because of the higher contribution of low-mass objects, inducing a deficit of predicted long duration events.
\subsection{Limitations of this study}
We have made a considerable effort to understand the CMDs and the microlensing data toward directions
that have not been examined by other teams. For this reason, we note the limits we encountered during this study to avoid any missinterpretation.
Knowledge of the absorption map was one of the most important limitations. Its precision and
resolution within the studied fields are parameters that impact the CMD so strongly that we found it necessary
to assume (reasonable) systematic and statistical dispersions to understand the observed densities of bright stars.
The blending and the 2:3 estimated fraction of binary stars are also other sources of limitation for understanding the CMDs.
All of these elements have fortunately a somewhat degenerated impact on the predicted stellar densities;
without any correction to the extinctions, we found that the simulated CMDs had too many stars
and were bluer than the data, which could be solved with a systematic extinction increase.
These limitations impacts mainly the CMDs; the specific observables from microlensing
(optical depth and durations) are mainly impacted through the distance distribution of the lenses.

\section{Conclusions and perspectives}
We have performed a complete simulation of the Galactic structure
and the EROS acceptance, which is able to reproduce all the EROS
exclusive observations toward the Galactic arms.
In this view, we produced a debiased color-magnitude diagram
from the HIPPARCOS catalog to feed our simulation with a
realistic stellar population.
This population was spatially distributed according to
the Besan\c{c}on Galactic model, and to a simple
Galactic mass model including a thin disk and a central bar, with an
adjustable thick disk contribution and IMF.
Every simulated object was then considered as a potential gravitational
lens as well as a potential source to gravitational lensing.
Taking into account the dust extinction and EROS detection efficiencies,
the observed color-magnitude diagrams and the microlensing optical depths and
durations are correctly fitted with both our simple Galactic model (with no
thick disk) and the Besan\c{c}on model.
We then used the simulation as a tool to obtain information on the
configuration space of the microlensing actors (lens and source
distance distributions).
The large width found in this way for the source distance distribution
validates the concept of ``catalog optical depth'' by
contrast with the usual optical depth to a given distance.
This concept is to be used as soon as the sources are
widely distributed in distance.
Finally, even with the small statistics of microlensing events, we
were able to extract interesting constraints on the Galactic
parameters -- {\it i.e.}, bar inclination confirmation, disk kinematics, mass function, and
hidden matter-- that have an impact on the microlensing
distributions.

The running VISTA Variables in the Via Lactea (VVV) survey, which
is monitoring stars within the Galactic plane in IR, is well suited to enlarge
the field of view within the Galactic plane, by searching for microlensing in dusty regions.
This survey should be able to better constrain the parameters mentioned above, with promising perpectives
such as measuring the mass function in areas other than the solar
neighborhood.
The Large Synoptic Survey Telescope (LSST) will also have the capability to monitor a wide domain
of the Galactic plane for microlensing, but only limited to the clear windows, free from large dust column densities.

\begin{acknowledgements}
This research was supported by the Perimeter Institute for 
Theoretical Physics and the John Templeton Foundation. Research at the Perimeter
Institute was supported by the Government of Canada through
Industry Canada and by the Province of Ontario through the Ministry of Economic 
Development and Innovation.
\end{acknowledgements}

\appendix

\section{Producing a local debiased CMD from the HIPPARCOS catalog}
\begin{figure}[htbp]
\begin{center}
\includegraphics[width=9.5cm]{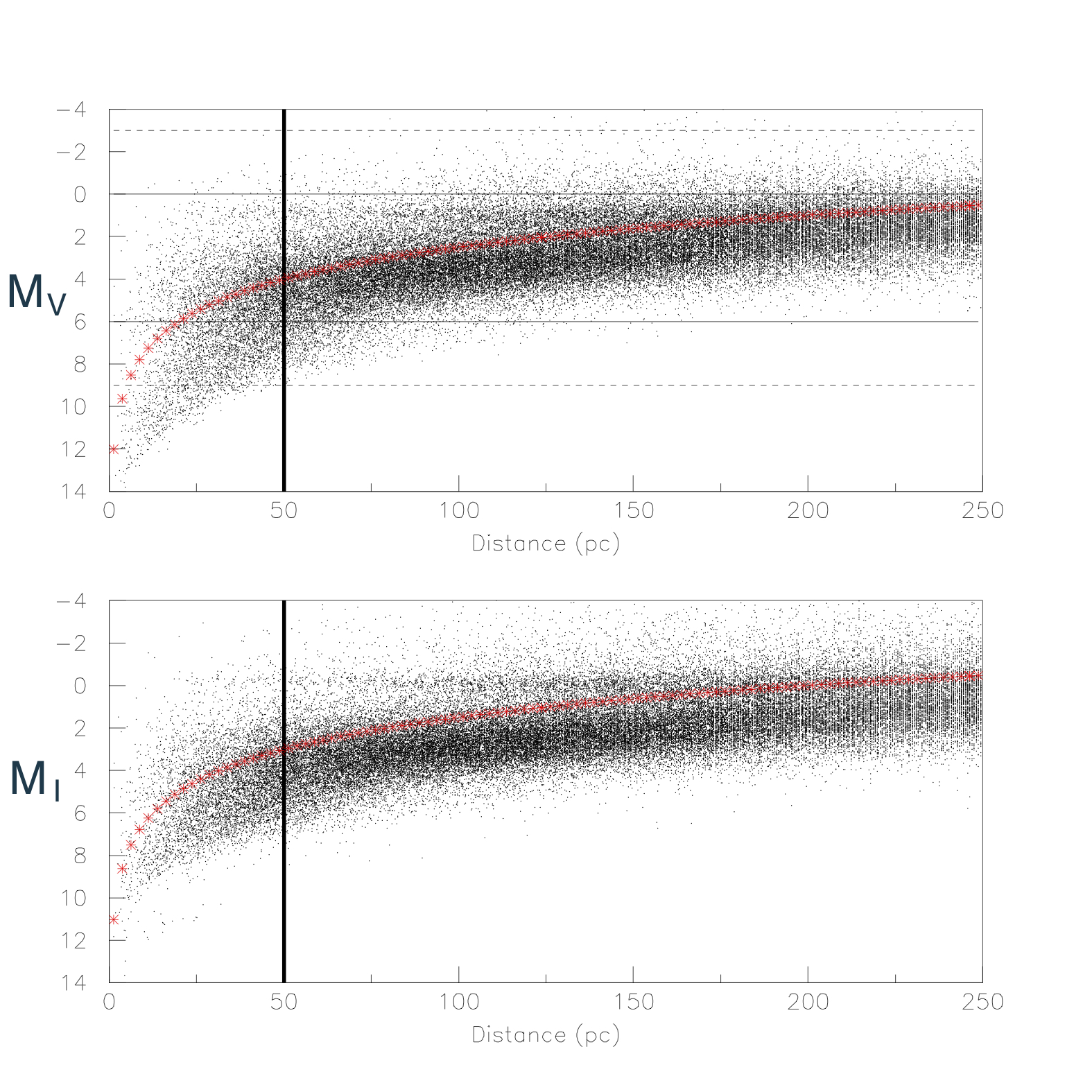}
\caption[]{
HIPPARCOS absolute magnitudes vs. distance distributions (up=$M_V$,
down=$M_I$). The red curves indicate
the absolute magnitude completeness limit as a function of the distance.
The vertical line shows our distance limit to get the local stellar
population.
The horizontal full lines at $M_V=0$ and $M_V=6$
correspond to the domain that contains enough stars from the HIPPARCOS
catalog to enable our debiasing procedure.
}
\label{mabs_dist}
\end{center}
\end{figure}

\begin{figure}[htbp]
\begin{center}
\includegraphics[width=8.5cm]{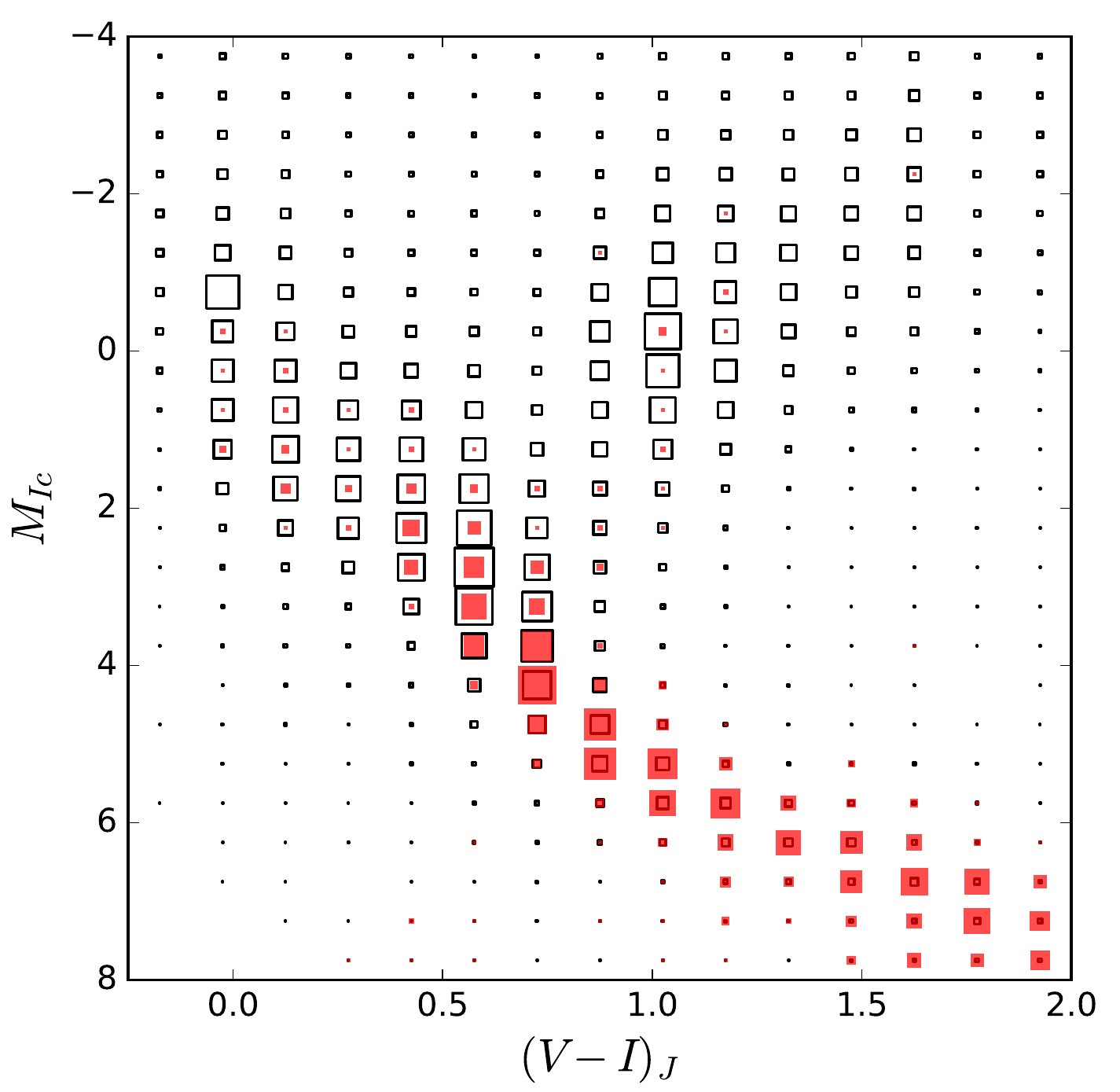}
\caption[]{
HIPPARCOS absolute color-magnitude diagram in $M_{I_{C}}$ vs.
$(V-I)_{J}$. The black squares correspond to the full catalog (statistically
biased).
The red squares correspond to the subsample of stars closer than $50 \mathrm{pc}$;
this subsample is statistically unbiased only for absolute magnitude
$M_V<4.0$ (corresponding to $M_I<3.1$,
above the horizontal line in the diagram).
The size scales are different between the red and black squares for readability.
}
\label{hr_all}
\end{center}
\end{figure}

The HIPPARCOS catalog provides equatorial coordinates $(\alpha, \delta)$,
apparent magnitudes $V_J$ ($=V$),
color indexes $(B-V)_J$, $(V-I)$, and parallaxes $\pi$.
To produce a local color-magnitude diagram, we calculate the absolute
magnitudes $M$ from the relative magnitudes and from the parallax, neglecting
the local absorption.
Fig. \ref{mabs_dist} shows the distribution of these absolute magnitudes $M_V$ and $M_I$
as a function of the distance for the catalogued stars.
It has been established (\cite{Jahreiss}) that the HIPPARCOS catalog is complete
until apparent visual magnitude $V=7.5$,
{\it i.e.}, above the red curves of fig. \ref{mabs_dist}.
This means that for a given absolute magnitude ${M_V}$,
the catalog is complete up to the distance $d_{c}(M_V)$ associated with the
distance modulus $\mu_{c}=7.5-M_V$;
for example, within $50\ pc$ the catalog is complete
up to $M_V=4.0$, which corresponds approximately to $M_I=3.1$.
Since we want to estimate the {\it local} CMD, we considered only those objects closer than $50 pc$
to avoid bias due to the very fast density variations with the distance to the Galactic plane.
Fig. \ref{hr_all} shows the full HIPPARCOS-Tycho $M_I$ versus $V-I$ 
distribution and the distribution limited to stars within $50 \mathrm{pc}$ (in red).
It is clear that the full catalog is strongly biased in favor
of bright (remote) objects.

\begin{figure}[htbp]
\begin{center}
\includegraphics[width=4.5cm]{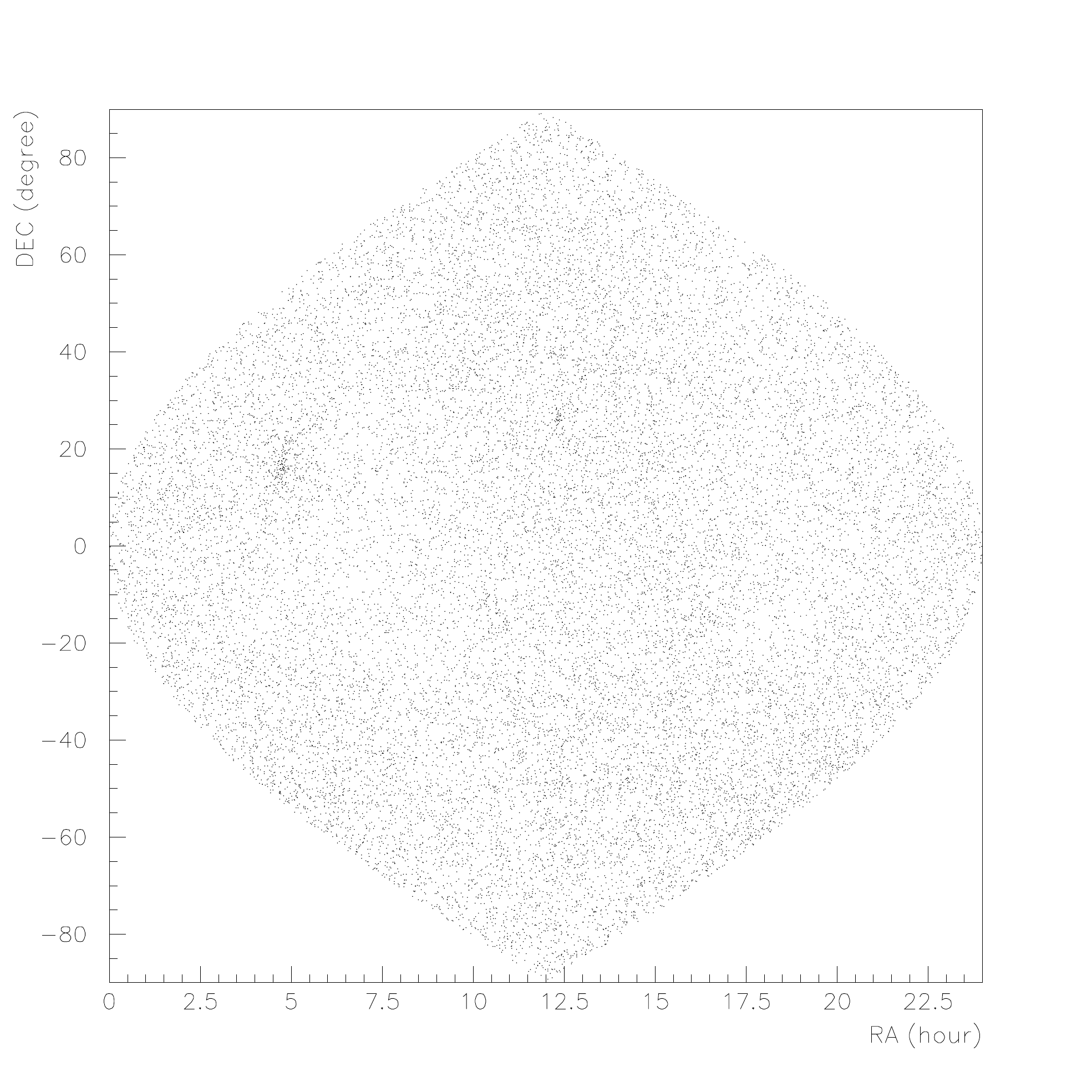}\includegraphics[width=4.5cm]{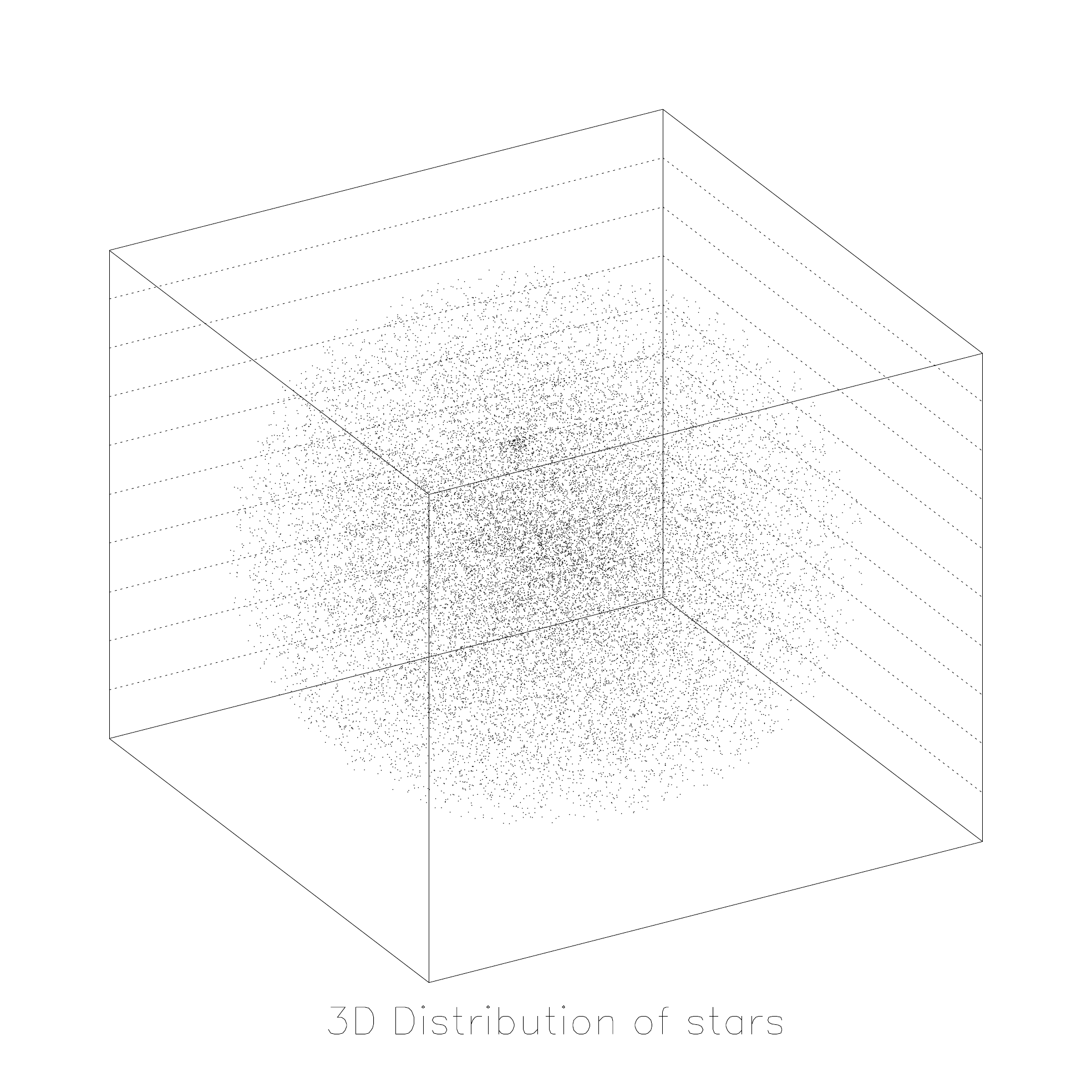}
\caption[]{
Two-dimensional and 3D distributions of the HIPPARCOS objects within $50\,\mathrm{pc}$.
The excess toward ($\alpha=67\degree\ \delta=16\degree$)
corresponds to the Hyades open cluster.
}
\label{map}
\end{center}
\end{figure}

To benefit from the whole statistics without suffering from selection bias,
we calculate the differential volumic density of stars as a function of the
absolute magnitude $0<M_V<6$ (interval chosen for statistical reasons, see next subsection)
from the numbers of stars found within the corresponding completion distance
\begin{equation}
d_{c}(M_V)=10\mathrm{pc} \times 10^{\frac{\mu_c}{5}}=10\mathrm{pc} \times 10^{\frac{7.5-M_V}{5}}
\simeq 50\mathrm{pc} \times 10^{\frac{4.0-M_V}{5}},
\label{dcomp}
\end{equation}
divided by the corresponding completion
volume $4\pi /3\times d_{c}(M_V) ^3$.
Those stars that we accounted for lie above the completion (red) curve
and between the two horizontal full lines in Fig. \ref{mabs_dist}.
As we need a diagram that is representative of the solar
neighborhood, we also consider only those stars that are inside a sphere
of radius $50 \mathrm{pc}$ (left of the vertical line in Fig. \ref{mabs_dist}) to avoid
depleted regions away from the Galactic median plane; indeed, as shown in
Fig. \ref{map}, the spatial 2D and 3D distributions of stars
within $50\,\mathrm{pc}$ distance of the catalog do not show global
anisotropies.
With all these constraints, a total of 2307 stars from the HIPPARCOS catalog
are used to build our debiased local CMD.
The upper panels of Fig. \ref{Hipp_50pc}  show the absolute magnitude
and color distributions of all the HIPPARCOS stars within $50 \mathrm{pc}$
(full lines) and of the stars that are closer than $min(d_{c}(M_V),50\,\mathrm{pc})$,
where $d_{c}(M_V)$ is the completion distance defined in Eq. (\ref{dcomp})
(dashed red lines).

We represent the HIPPARCOS catalog as a multi-dimensional distribution function
defined by
\begin{equation}
f\left( {\bf x},{\bf M} \right)
=\sum_{catalog} {\delta}({\bf M}-{\bf M_{i}}) {\delta}^3 ({\bf x}-{\bf x_{i}}),
\end{equation}
where ${\bf x_{i}}$ is the position of star $i$ and ${\bf M_{i}}$ represents
its absolute magnitude and color ``vector'' ({\it i.e.}, its type).
As explained above, to extract the unbiased local density
for a given stellar type characterized by the vector ${\bf M}$ (here $(M_I,M_V)$), we
only account for the objects that are both within the completion
volume ($d<d_{c}(M_V)$) and closer than $50\, \mathrm{pc}$, {\it i.e.},
\begin{equation}
n({\bf M})= \frac{3}{4\pi .min[{d_{c}(M_V)}, 50\mathrm{pc}]^3}\!\!
\int_{d<min[{d_{c}(M_V)}, 50\mathrm{pc}]} \!\!\! f( {\bf x},{\bf M}) k(d) d^3 x,
\end{equation}
where $k(d)$ is a correction factor that takes into account the
variation of the density within the completion volume (this
correction varies from $1$ to $1.09$).

\subsection{ Extrapolating the local HIPPARCOS CMD}
The number of usable HIPPARCOS objects (closer than $min(d_{c}(M_V),50\,\mathrm{pc})$)
is statistically limited in the faint ($M_V>6$) and bright ($M_V<0$) ends,
as can be seen in Fig. \ref{Hipp_50pc} (upper left, dashed line).
Moreover, there is no star with $M_V>9$ within its corresponding completion
distance $d_{c}(9)\simeq 5\mathrm{pc}$ ({\it i.e.}, above the red curve of Fig. \ref{mabs_dist}),
because the volume is too small;
there is also no local star (within $50\,\mathrm{pc}$) brighter than $M_V=-3$.

Therefore, when building a debiased density color-magnitude diagram,
we need to examine specifically the contribution of
the stars with absolute $M_V$ magnitudes out of $[0 ,6]$ range
to avoid statistical limitations or biases:
\begin{figure}[htbp]
\begin{center}
\includegraphics[width=9.cm]{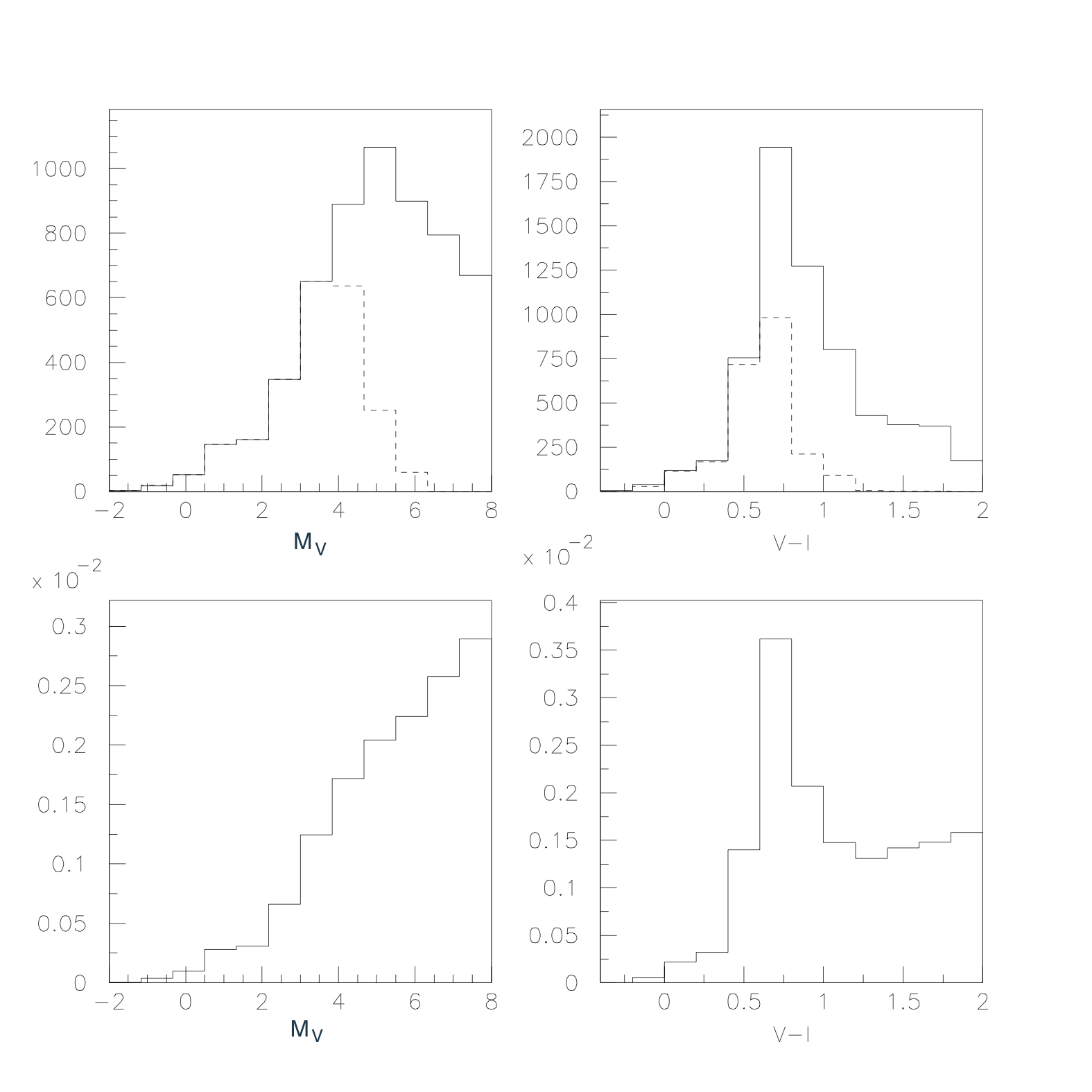}
\includegraphics[width=9.cm]{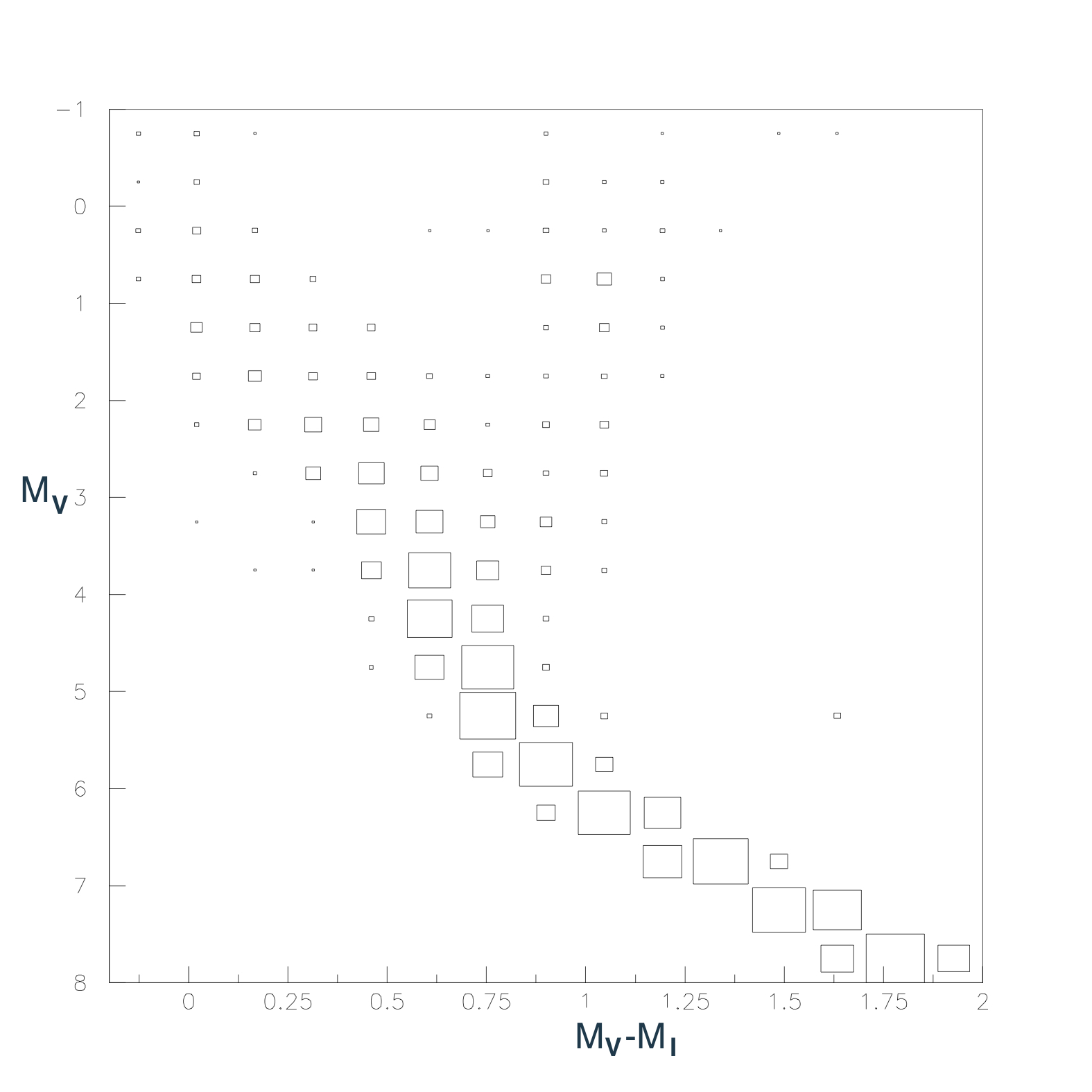}
\caption[]{
Top: Raw distributions of $M_V$ and $(V-I)$ of all HIPPARCOS stars
within $50 \mathrm{pc}$ (6911 objects).
The dashed lines show the numbers of stars within
the completion volume corresponding to their magnitude (see text).
Middle: Local debiased volumic density of stars
(per magnitude unit, in $\mathrm{\mathrm{pc^{-3}}}$)
estimated from the ratio of stars within the
completion volume and extrapolated beyond $M_V=6$.
Bottom: Debiased $M_V$ vs. $V-I$ stellar density of stars
closer than $50\, \mathrm{pc}$. 
}
\label{Hipp_50pc}
\end{center}
\end{figure}
\begin{itemize}
\item
First, we can neglect the contribution of the brightest stars; indeed, the
HIPPARCOS catalog contains only 35 stars brighter than $M_V<0$ within $50 \mathrm{pc}$
(complete sample).
This corresponds to a maximum contribution of
\begin{equation}
35\times\left[\frac{10\mathrm{kpc}}{50\mathrm{pc}}\right]^3\times\frac{\Omega (1\degree\!\!\times\! 1\degree)}{4\pi}\sim 6800\, {\rm stars/sq.\, deg.}
\end{equation}
of $M_V<0$ stars within $10 \mathrm{kpc}$ distance (typically less than $2-3\%$
toward the directions studied in this paper). This contribution will be neglected in the following discussions
\footnote{
For this very conservative estimate, we assume a constant density along the line of
sight, we neglect the absorption, and we assume a $100\%$ detection efficiency.}.
\item
Stars fainter than $M_V=6$ have a minor, but not negligible contribution to a deep Galatic exposure.
Instead of debiasing the statistically limited subsample of the HIPPARCOS catalog, we choose to linearly
extrapolate the local stellar density of these faint stars as (see Fig. \ref{Hipp_50pc} middle, left):
\begin{equation}
\frac{\ud n}{\ud M_V}=Const.+4.6\times 10^{-4} M_V\ (\mathrm{pc^{-3}mag^{-1}}).
\end{equation}
Since we deduce from Fig. \ref{hr_all} (lower right branch) that
\begin{equation}
V-I\sim 0.47\times (M_I-3.97), 
\end{equation}
or equivalently 
\begin{equation}
 V-I\sim 0.33\times (M_V-4.0) 
\label{colmag}
\end{equation}
the type of these faint stars is also completely extrapolated.
\end{itemize}
Fig. \ref{Hipp_50pc} (bottom) shows the local CMD obtained
following our complete procedure using the HIPPARCOS stars
with $0<M_V<6$ within $d_{c}(M_V)$ and our extrapolated distribution
for $6<M_V<8$.
\subsection{Comparison with the stellar density expected from the mass
function: A coherence check}
\label{section:check}
We can crosscheck the stellar number density found from
the HIPPARCOS catalog and the density expected from the mass
function as follows:
stars with $0<M_V<6$ belong to the mass domain defined by
$0.85\Msol <m<2.8\Msol$  (\cite{Delfosse}).
The local number density of objects within this mass range 
is given by
\begin{equation}
n(0.85\Msol<m<2.8\Msol)=
\int_{0.85}^{2.8}{\frac{\ud n}{\ud m}\ud m},
\end{equation}
where $\frac{\ud n}{\ud m}$ is the stellar mass function in the solar neighborhood.
We use the mass function $\xi(\log m/\Msol)=\frac{\ud n}{\ud \log m/\Msol}$
of \cite{Chabrier 2003}, revised in \cite{Chabrier 2004},
\begin{eqnarray}
\xi(\log m/\Msol)&=&
0.093\times exp\left[ \frac{-(\log m/0.2\Msol)^2}{2\times (0.55)^2}\right],\,m\le \Msol \nonumber \\
&=&0.041 (m/\Msol)^{-1.35},\,m> \Msol
\label{chabrierfunction}
\end{eqnarray}
(see fig \ref{fitdyn}).
We find that the mean density of disk stars with $0<M_V<6$
in a sphere of $50\, \mathrm{pc}$ centered on the sun (located at $26\, \mathrm{pc}$ from the
disk plane (\cite{Majaess})) is $0.012 \mathrm{pc^{-3}}$. 
This is compatible with the estimates from the integral of the
$M_V$ debiased distribution of the volumic density of stars plotted
in Fig. \ref{Hipp_50pc}, $n_{HIPPARCOS}=0.0076 \mathrm{pc^{-3}}$, when taking
into account the fact that $\sim$ 2:3 of the stars are in binary
systems (\cite{Chabrier 2004}) not deblended in the HIPPARCOS
observations.

\section{Parameters of the Besan\c{c}on Galactic model}
The Sun is located at $\mathrm{R_{\odot}=8.0 kpc}$ and $\mathrm{z_{\odot}=15\ pc}$, which is different than in our simple model.
The thin disk structures are parametrized in cylindrical galactocentric coordinates $(r,z)$, and for various ranges of age, as follows:
\begin{eqnarray}
\rho_{D}(r,z)_{age}  &\propto& \left[ \exp(-\frac{a^2}{R_d^2} ) -\exp(-\frac{a^2}{R_h^2} ) \right]~\mathrm{if~age<0.15~Gyr}, \nonumber \\
   &\propto& \left[ \exp\left(- \sqrt{0.25 + \frac{a^2}{R_d^2}}\right) -\exp\left(-\sqrt{0.25 +\frac{a^2}{R_h^2}}\right) \right], \nonumber \\
&&\mathrm{if~age>0.15~Gyr,}
\end{eqnarray}
where
\begin{itemize}
\item
$R_d = 5.0$ kpc and  $R_h = 3.0$ kpc if age$<0.15$ Gyr,
\item
$R_d = 2.17$ kpc and  $R_h = 1.33$ kpc if age$>0.15$ Gyr,
\item
$a^2 = r^2 + (z/\epsilon_{age})^2$;
\item
$\epsilon_{age}$ and the local mass densities corresponding to $\rho_{D}(r_{\odot},z_{\odot})_{age}$
values are given in table \ref{local_density} for the different ranges of stellar age, together with the IMFs.
\end{itemize}

The thick disk contribution is expressed by
\begin{eqnarray}
\rho_{D}^{thick}(r,z)  &=& \rho_{D}^{thick}(r_{\odot},z_{\odot}) \\
			& & \times (1-\frac{z^2}{x_l (2h_z+x_l)}) \exp\left[-\frac{r-R_{\odot}}{R_{thick}} \right]~\mathrm{if}~|z|<x_l, \nonumber \\
   & & \times \frac{\exp(x_l/h_z)}{1+x_l/2h_z}\exp\left[-\frac{|z|}{h_z}\right] \exp \left[ -\frac{r-R_{\odot}}{R_{thick}} \right]~
\mathrm{if}~|z|>x_l, \nonumber
\end{eqnarray}
where $x_l=\mathrm{400pc}$, $h_z=\mathrm{800pc}$ and $R_{thick}=\mathrm{2.5Kpc}$.
Table \ref{local_density} also gives the total local densitiy $\rho_{D}^{thick}(r_{\odot},z_{\odot})$
for the thick disk together with the IMF.

The two components of the bar are described in
a Cartesian frame positioned at the Galactic center with the major
axis X tilted by $\Phi= 12.8$ degree with respect to the
Galactic center-Sun direction.
The mass density
for each component of the bar is given by (\cite{Robin})
\begin{equation}
\rho_{bar1}(X,Y,Z)= \rho_0 sech^2(-R_s)\times f_c(X,Y)
\end{equation}
\begin{equation}
\rho_{bar2}(X,Y,Z)= \rho_0 \exp(-R_s)\times f_c(X,Y),
\end{equation}
where
\begin{equation}
R_s^{C_p}= \left[\left| \frac{X}{a}\right|^{C_n} + \left| \frac{Y}{b}\right|^{C_n} \right]^{\frac{C_p}{C_n}} + \left| \frac{Z}{c}\right|^{C_p},  \\
\end{equation}
and $f_c$ is a cutoff function
\begin{eqnarray}
f_c(X,Y)&=&1.~~~\mathrm{if}~X^2+Y^2<R_c^2,  \\
 &=&exp\left[ -\frac{(\sqrt{X^2+Y^2}-R_C)^2}{0.25\,\rm{kpc}^2}\right]~\mathrm{if}~X^2+Y^2>R_c^2. \nonumber
\end{eqnarray}
The parameters for bar1 are $\rho_0=9.21\, \mathrm{\Msol.pc^{-3}}$
(\footnote{Not to be confused with the local density of the bar $\rho_{bar}(r_{\odot},z_{\odot})$}),
$a=1.46$ kpc, $b = 0.49$ kpc, $c=0.39$~kpc, $R_c=3.43$ kpc are the scale length factors and
$C_p=3.007$, $C_n=3.329$. The total mass of this bar is $35.45\times 10^{9}\mathrm{\Msol}$\\
The parameters for bar2 are $\rho_0= 0.026\, \mathrm{\Msol.pc^{-3}}$, $a=4.44$ kpc,
$b = 1.31$ kpc, $c=0.80$~kpc, $R_c=6.83$ kpc are the scale length factors, and
$C_p=2.786$, $C_n=3.917$. The total mass of this bar is $2.27\times 10^{9}\mathrm{\Msol}$\\

The IMF for these two bars is dn/dm $\propto$ $(m/\Msol)^{-2.35}$.

\begin{table}[h]
{\centering
\caption{Age, local mass density $\rho(r_{\odot},z_{\odot})$, disk axis ratio $\epsilon$, and IMF
of the different stellar components of the disks in the Besan\c{c}on model.
WD represents the white dwarfs.
}
\label{local_density}
\begin{tabular}{lllll}
\hline
        &Age    &$\rho(r_{\odot},z_{\odot})$   &$\epsilon$ & IMF\\
        &(Gyr)  &($\Msol \mathrm{pc}^{-3}$)  &\\
\hline
disk    &0-0.15 &4.0$\times$10$^{-3}$    &0.0140 & \\
        &0.15-1 &7.9$\times$10$^{-3}$       &0.0268 & \\
        &1-2    &6.2$\times$10$^{-3}$         &0.0375 & dn/dm $\propto (m/\Msol)^{-\alpha}$ \\
        &2-3    &4.0$\times$10$^{-3}$         &0.0551 & $\alpha$ = 1.6 for m $< 1 \Msol$ \\
        &3-5    &5.8$\times$10$^{-3}$         &0.0696 & $\alpha$ = 3.0 for m $> 1 \Msol$ \\
        &5-7    &4.9$\times$10$^{-3}$         &0.0785 & \\
        &7-10   &6.6$\times$10$^{-3}$        &0.0791 & \\
  &WD & 3.96$\times$10$^{-3}$   \\
\\
Thick disk & all &1.64$\times$10$^{-3}$  &               & dn/dm $\propto (m/\Msol)^{-0.5}$ \\
\hline
\end{tabular}\par}
\end{table}

As far as kinematics is concerned,
we use the ellipsoids of velocity dispersions provided for each
structure and age in table 4 of (\cite{besancon}).

\end{document}